%% file: main.tex
\documentclass[superscriptaddress,
nofootinbib,
nobibnotes,
notitlepage,
amsmath,amssymb,
aps,
twocolumn,
prd,
]{revtex4-2}

\usepackage{bm}
\usepackage{color}
\usepackage{xspace}
\usepackage{multirow}
\usepackage{rotating}
\usepackage{threeparttable}
\usepackage{tabularx}
\usepackage{hyperref}

\def\beq{\begin{eqnarray}}
\def\eeq{\end{eqnarray}}

\usepackage[normalem]{ulem}
\usepackage{physics}
\usepackage{ifthen}
\usepackage{aas_macros}
\usepackage{dcolumn}
\newcolumntype{d}{D{.}{.}{-1}}
\newcolumntype{C}{>{$}c<{$}}
\graphicspath{{}}

\newcommand{\Mpch}{h^{-1}\mathrm{Mpc}}
\newcommand{\hMpc}{h\,\mathrm{Mpc}^{-1}}
\newcommand{\hun}{\,\mathrm{km}\,\mathrm{s}^{-1}\mathrm{Mpc}^{-1}}

\newcommand{\editA}[1]{#1}
\newcommand{\editB}[1]{#1}
\newcommand{\editC}[1]{#1}
\newcommand{\resub}[1]{#1}

\newcommand*{\specialcell}[2][l]{%
  \begin{tabular}[c]{@{}#1@{}}#2\end{tabular}}

\newenvironment{rcases}
  {\left.\begin{aligned}}
  {\end{aligned}\right\rbrace}

\begin{document}

%\title{Cosmology with unWISE and CMB lensing tomography from ACT}
\title{The Atacama Cosmology Telescope: Multi-probe cosmology with unWISE galaxies and ACT DR6 CMB lensing}
%\shorttitle{unWISE x ACT DR6 3$\times$2pt cosmology}
% \shortauthors{Farren, Krolewski, Qu, Ferraro, Sherwin et al.}

\input{authors_farren.tex}

\date{\today}% It is always \today, today,

\begin{abstract} 
We present a joint analysis of the CMB lensing power spectra measured from the Data Release 6 of the Atacama Cosmology Telescope and \textit{Planck} PR4, cross-correlations between the ACT and \textit{Planck} lensing reconstruction and galaxy clustering from unWISE, and the unWISE clustering auto-spectrum. We obtain $1.5\%$ constraints on the matter density fluctuations at late times parametrised by the best constrained parameter combination $S_8^{\rm 3x2pt}\equiv \sigma_8 (\Omega_m/0.3)^{0.4} = 0.815 \pm 0.012$. The commonly used $S_8 \equiv \sigma_8 (\Omega_m/0.3)^{0.5}$ parameter is constrained to $S_8 = 0.816\pm0.015$. In combination with baryon acoustic oscillation (BAO) measurements we find $\sigma_8=0.815\pm 0.012$. We also present sound-horizon-independent estimates of the present day Hubble rate of $H_0=66.4^{+3.2}_{-3.7} \hun$ from our large scale structure data alone and $H_0=64.3^{+2.1}_{-2.4} \hun$ in combination with uncalibrated supernovae from \texttt{Pantheon+}. \editB{Using parametric estimates of the evolution of matter density fluctuations, we place constraints on cosmic structure in a range of high redshifts typically inaccessible with cross-correlation analyses.} Combining lensing cross- and auto-correlations, we derive a 3.3\% constraint on the integrated matter density fluctuations above $z=2.4$, one of the tightest constraints in this redshift range and fully consistent with a $\Lambda$CDM model fit to the primary CMB from \textit{Planck}. Finally, combining with primary CMB observations and using the extended low redshift coverage of these combined data sets we derive constraints on a variety of extensions to the $\Lambda$CDM model including massive neutrinos, spatial curvature, and dark energy. We find in flat $\Lambda$CDM $\sum m_\nu<0.12$ eV at 95\% confidence using the LSS data, BAO measurements from SDSS and primary CMB observations.
\end{abstract}

\maketitle
%\tableofcontents

\section{Introduction}\label{sec: intro}

Measurements of the matter density fluctuations at low redshifts inform our understanding of the formation of cosmic structure, probe the nature of dark matter and dark energy, and constrain the masses of neutrinos. They also provide an important test of the predictions of general relativity. Gravitational lensing observations that are sensitive to the total matter distribution, including the invisible dark matter, have become an indispensable tool for studying cosmic structure. Several lensing related techniques have been developed to study both the weak gravitational lensing of galaxies as well as of the cosmic microwave background (CMB).

Over the past two decades a standard model of cosmology has emerged primarily based on high precision observations of the CMB. Measurements by \textit{WMAP} first established the now prevailing six parameter $\Lambda$CDM model \citep{2003ApJS..148..175S,2013ApJS..208...19H}. It posits that the Universe is dominated by phenomenological cold dark matter (CDM), is spatially flat, and its expansion is driven by a cosmological constant $\Lambda$. These results were sharpened by measurements made by the \textit{Planck} satellite \citep{2020A+A...641A...6P}. The model also makes predictions for other cosmological observables, which in recent years have reached increasing precision enabling new tests of this model. Despite the $\Lambda$CDM model's overall success, some discrepancies have been observed and several extensions have been put forward. Here we weigh in on some of these discrepancies and constrain extensions beyond the standard model.

\editB{One such discrepancy and a primary focus of this work is the amplitude of matter density fluctuations typically parametrised in terms of $\sigma_8$, the RMS (root-mean-square) of the linear matter density contrast smoothed on scales of $8 \Mpch$, which provides the normalisation of the matter power spectrum. The shape and redshift evolution of the matter power spectrum are predicted from the $\Lambda$CDM model.}

In recent work \cite{2024ApJ...962..112Q} and \cite{2024ApJ...962..113M} reconstructed the gravitational lensing field over 9400 deg$^2$ from new high resolution CMB observations by the Atacama Cosmology Telescope (ACT). They showed percent level constraints on the integrated matter density fluctuations over a wide range of redshifts ($z\lesssim 5$) from the CMB lensing power spectrum. The $\sigma_8$ constraints are in excellent agreement with model extrapolations from a $\Lambda$CDM model fit to observations of the primary fluctuations in the CMB as observed by \textit{Planck} \citep{2020A+A...641A...6P}. Their results are also in excellent agreement with CMB lensing measurements from \textit{Planck} \citep{2020A+A...641A...8P, 2022JCAP...09..039C} and the combination of both measurements yields improved constraints on cosmic structure.

\cite{2024ApJ...966..157F} used the ACT DR6 and \textit{Planck} CMB lensing reconstruction together with galaxies detected in imaging data from the Wide-Field Infrared Survey Explorer \citep[WISE;][]{2010AJ....140.1868W} to focus on a lower redshift range, approximately $0.2\lesssim z \lesssim 1.8$. Using the correlation between the galaxy distribution in two redshift bins, which acts as a biased tracer of the underlying matter density, and the CMB lensing reconstruction, \cite{2024ApJ...966..157F} similarly found good agreement with the $\Lambda$CDM prediction for $\sigma_8$ and CMB lensing auto-correlation results.

Meanwhile, several galaxy weak lensing surveys like the Dark Energy Survey \citep[DES;][]{2015AJ....150..150F,2022PhRvD.105b3520A}, the Kilo-Degree Survey \cite[KiDS;][]{2015MNRAS.454.3500K,2021A+A...646A.140H}, and the Hyper Suprime-Cam \cite[HSC;][]{2018PASJ...70S...4A,2023PhRvD.108l3520M,2023PhRvD.108l3517M,2023PhRvD.108l3521S}, among others, have found a $2-3\sigma$ lower amplitude of matter density fluctuations, compared to the prediction from \textit{Planck} primary CMB assuming a standard $\Lambda$CDM model. Such surveys typically probe lower redshifts, $z \lesssim 1$, than the CMB lensing work discussed above. Similar results have also been obtained by some other \editA{studies of CMB lensing cross-correlations with galaxy surveys}, albeit at varying levels of significance \citep[see e.g.,][]{2020MNRAS.491...51S,2021MNRAS.501.6181K,2021MNRAS.501.1481H, 2021A+A...649A.146R,2022JCAP...02..007W,2022JCAP...07..041C,2021JCAP...12..028K,2023PhRvD.107b3530C,2023PhRvD.107b3531A,2024JCAP...01..033M,Kim2024,Sailer2024}. \editA{While the modelling assumptions and the range of scales probed in these works vary, they mostly sensitive to lower redshifts than the analysis presented in \cite{2024ApJ...966..157F}.}

This motivates a further investigation of the formation of structure at low redshifts. We note that the  $\sigma_8$ constraints from \cite{2024ApJ...962..112Q, 2024ApJ...962..113M} and \cite{2024ApJ...966..157F} differ from the galaxy weak lensing result not only in terms of the redshift range, but also in the scales probed. As pointed out for example by \cite{2022MNRAS.516.5355A} and \cite{2023MNRAS.525.5554P} galaxy weak lensing draws significant information from highly non-linear scales which the CMB lensing auto- and cross-correlations are insensitive to. The observed discrepancy \editA{($\sim$$2-3\sigma$)} with the $\Lambda$CDM prediction for $\sigma_8$ from \textit{Planck} may therefore also be explained by a suppression of the matter power spectrum on small scales that exceeds expectations of the baryon feedback induced suppression based on \editA{the most recent} hydrodynamical simulations \citep{2022MNRAS.516.5355A,2023MNRAS.525.5554P,2023MNRAS.526.5494M}.

Motivated in part by further investigation of the formation of structure at low redshifts and on linear to mildly non-linear scales we combine the lensing auto-spectrum analysis from \cite{2024ApJ...962..112Q} and \cite{2024ApJ...962..113M} with the cross-correlation analysis from \cite{2024ApJ...966..157F}. Throughout the paper we will often refer to this combination as `3x2pt' given that it contains three two-point correlation functions (or rather their harmonic space equivalent, the power spectrum). These are the auto-spectrum of the CMB lensing convergence, $C_\ell^{\kappa \kappa}$, the cross-correlation between galaxies and CMB lensing, $C_\ell^{\kappa  g}$, and the galaxy auto-correlation, $C_\ell^{gg}$. By contrast we will often refer to the cross-correlation analysis as `2x2pt' ($C_\ell^{\kappa  g}$ \& $C_\ell^{gg}$). In Sec.\,\ref{subsubsec:structure_growth} we show constraints on cosmic structure formation from our `3x2pt' data.

Another discrepancy, which has reached $\sim$$5\sigma$ with recent data, is between the present day expansion rate, parameterised by the Hubble constant $H_0$, inferred within the $\Lambda$CDM model from the CMB \citep{2020A+A...641A...6P} and a local measurement based on Cepheid-calibrated supernovae from \texttt{SH0ES} \citep{2022ApJ...934L...7R,2024arXiv240408038B}. Meanwhile results from baryon acoustic oscillation (BAO) observations have generally found results consistent with the CMB-derived values \citep[e.g.,][]{2021PhRvD.103h3533A,2024arXiv240403002D}. CMB and BAO constraints are predominantly informed by the angular size of the sound horizon scale\footnote{Here, we do not make a careful distinction between the sound horizon scale relevant for BAO ($r_{\rm d}$) and CMB ($r_{\rm s}$) observations, although to be precise these are defined at the baryon drag epoch and at photon decoupling, respectively.}. This fact has motivated theoretical work to explain the tension by invoking new physics that decreases the physical size of the sound horizon at recombination by approximately $10\%$ (e.g.,~\citealp{2019ApJ...874....4A,2020PhRvD.101d3533K}). It also motivates new measurements of the Hubble constant that are derived from a different physical scale present in the large-scale structure, namely, the matter-radiation equality scale (with comoving wave-number $k_{\rm eq}$) which sets the turn-over in the matter power spectrum. As pointed out in \cite{2021MNRAS.501.1823B} such measurements can be obtained from the CMB lensing power spectrum and other large scale structure (LSS) tracers and we investigate the implications of our data for the $H_0$-tension in Sec.\,\ref{subsubsec:hubble}.

Beyond modifications to the flat $\Lambda$CDM model motivated by these observed discrepancies other extensions are motivated by physical considerations. Given the phenomenological nature of dark energy it is natural to consider departures from the cosmological constant model which allow an equation of state $w\neq -1$ or evolution in the dark energy equation of state. We consider such models in Sec.\,\ref{subsubsec:de_extensions}; our data unfortunately only marginally improves on existing constraints on such models derived from the primary CMB, BAO and supernovae. Furthermore, observations of neutrino oscillations require neutrinos to be massive, and while the \editA{mass splitting} between the neutrino mass eigenstates is well determined the absolute mass scale is poorly constrained \citep[see e.g.,][]{2022PrPNP.12403947A}. In Sec.\,\ref{subsubsec:nulcdm} we derive constraints on the neutrino mass from the characteristic suppression of the formation of structure on scales smaller than the neutrino free streaming scale which can be probed with our `3x2pt' data. It has been pointed out that those neutrino mass constraints are model dependent, as the effect of massive neutrinos can be mimicked also by some beyond-$\Lambda$CDM models \citep[see e.g.,][]{2015PhRvD..92l3535A,2018JCAP...09..017C,2019EPJC...79..262R,2020JCAP...07..037C}. Thus we consider massive neutrinos in the context of extended dark energy models in Sec.\,\ref{subsubsec:nu_de_extensions}. In addition to dark energy and neutrinos, we also revisit the assumption of spatial flatness in Sec.\,\ref{subsubsec:curvature} where our data provides competitive cross-checks on constraints from the primary CMB and BAO.

Before presenting these results in Sec.\,\ref{sec:cosmo} we discuss the data used in this analysis in Sec.\,\ref{sec:data} including the external data sets we employ in our analysis (Sec.\,\ref{subsec:external_data}). In Sec.\,\ref{sec:covmat} we briefly describe how we obtain the covariance for the likelihood analysis described in Sec.\,\ref{sec:analysis}. Finally, in Sec.\,\ref{sec:conclusion} we summarise our findings and conclude with an outlook to future work. For the convenience of the reader we also provide an overview of the key results from this work in the following section (Sec.\,\ref{sec:key_results}).

\section{Key Results}\label{sec:key_results}

\begin{figure*}
    \centering
    \includegraphics[width=\linewidth, trim=0cm 0.2cm 0cm 0.5cm, clip]{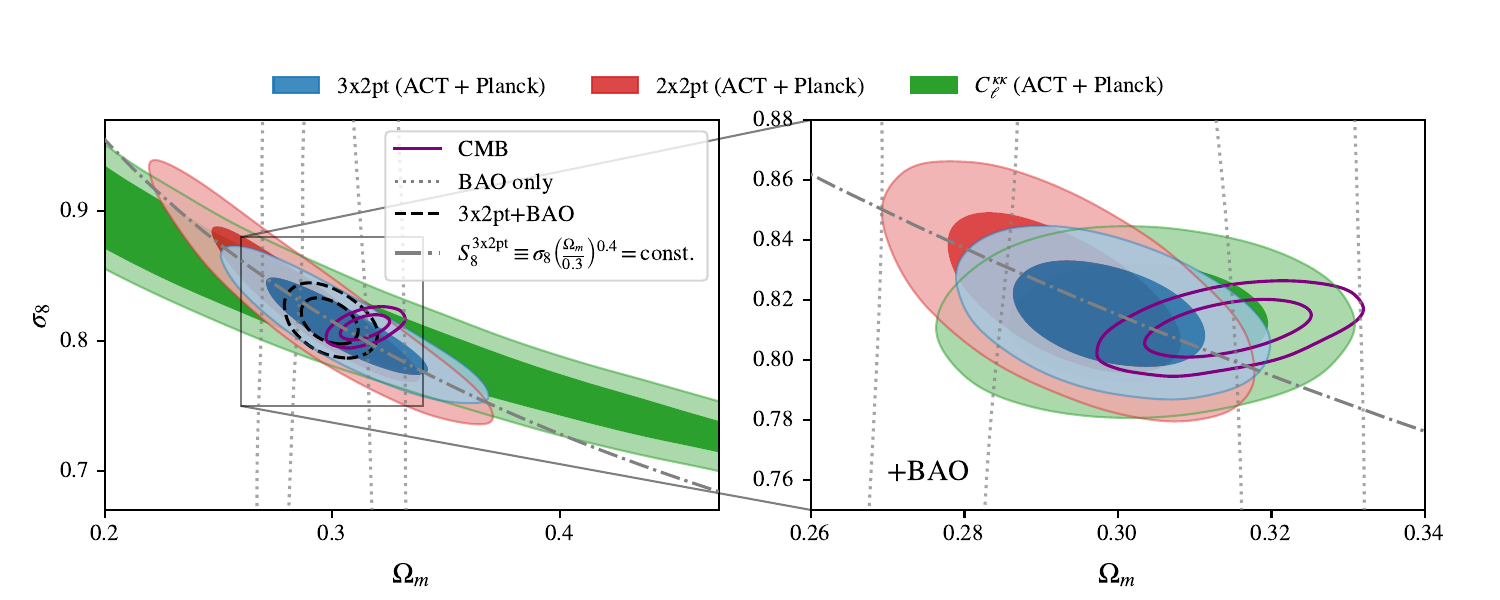}
    \caption{The key result from our analysis are tight constraints on the amplitude of matter density fluctuations. The left panel shows a comparison of constraints derived from the CMB lensing auto-correlation ($C_\ell^{\kappa \kappa}$), from the cross-correlation of CMB lensing and unWISE galaxies together with the unWISE auto-correlation (2x2pt), and from the combination of all three two-point functions (3x2pt). As is usual in these type of analyses the matter density fluctuations parametrised by $\sigma_8$ and the global matter density $\Omega_m$ are degenerate. The best constrained parameter combination for the 3x2pt analysis is $S_8^{\rm 3x2pt} \equiv \sigma_8 (\Omega_m/0.3)^{0.4}$ (indicated by a dash-dotted line in the left panel). In the dashed contours on the left \editB{(3x2pt+BAO)} and in the zoom in on the right we show the combination with BAO data which breaks the $\sigma_8$-$\Omega_m$ degeneracy. As can be seen by comparison to contours obtained from model predictions within a $\Lambda$CDM model fit to the primary CMB our results are in good agreement with CMB constraints from \textit{Planck} (solid purple, unfilled lines). All constraints shown here include lensing data from both ACT DR6 and \textit{Planck} PR4.}
    \label{fig:omegam_sigma8_3x2pt}
\end{figure*}

The key results from this work are constraints on the amplitude of matter density fluctuations at low redshift parameterised by $\sigma_8$. As in CMB lensing autocorrelation analyses \citep[e.g.,][]{2024ApJ...962..112Q,2022JCAP...09..039C}, CMB lensing cross-correlations \citep[e.g.,][]{2021MNRAS.501.1481H,2024ApJ...966..157F,Kim2024}, or galaxy weak lensing \citep[e.g.,][]{2022PhRvD.105b3520A,2021A+A...646A.140H,2023PhRvD.108l3517M,2023PhRvD.108l3521S}, when using the projected large scale structure tracers only, $\sigma_8$ is significantly degenerate with the matter density, $\Omega_m$ (as can be seen in the left panel of Fig.\,\ref{fig:omegam_sigma8_3x2pt}). In our case the parameter combination best constrained by the combination of CMB lensing auto-spectrum, galaxy-galaxy clustering auto-power spectra, and the cross-correlation between CMB lensing and galaxies is approximately $\sigma_8 \Omega_m^{0.4}$. In analogy with the commonly used parameter $S_8 \equiv \sigma_8 (\Omega_m/0.3)^{0.5}$, corresponding to the parameter combination best constrained by galaxy weak lensing surveys, we therefore define the parameter $S_8^{\rm 3x2pt} \equiv \sigma_8 (\Omega_m/0.3)^{0.4}$. 

Using the lensing auto-spectra from ACT and \textit{Planck}, the respective cross-correlations with the unWISE galaxies, and the galaxy auto-correlation (3x2pt) we constrain this parameter combination to $\sim$$1.5\%$,
\beq S_8^{\rm 3x2pt} \equiv& \sigma_8 (\Omega_m/0.3)^{0.4} = 0.815\pm 0.012\hspace{0.3cm} ({\rm 3x2pt}).\eeq
Within the $\Lambda$CDM model this parameter is similarly well constrained by primary CMB data ($S_8^{\rm 3x2pt} = 0.826\pm 0.012$ from the primary CMB data sets discussed in Sec.\,\ref{subsec:primary_CMB_data}) and the two constraints are in good agreement as can be seen in the left panel of Fig.\,\ref{fig:omegam_sigma8_3x2pt}. Even though $S_8$ differs slightly from the best constrained combination of $\sigma_8$ and $\Omega_m$ in our work we nevertheless also obtain highly competitive constraints on this parameter combination which can be compared, for example, to results from DES, KiDS and HSC \citep{2022PhRvD.105b3520A,2021A+A...646A.140H,2023PhRvD.108l3521S}. Using again both ACT and \textit{Planck} CMB lensing auto-spectra, the cross-correlations with unWISE, and the unWISE auto-spectrum we find \beq S_8=0.816\pm 0.015 \hspace{0.3cm} ({\rm 3x2pt}).\eeq While our value of $S_8$ falls between the typical values preferred by galaxy weak lensing analyses and those predicted from $\Lambda$CDM fits to the CMB, it is not in statistically significant tension with either of those data sets \editB{($\sim$$1\sigma$ and $\sim$$1.5-2.5\sigma$ relative to primary CMB and various galaxy weak lensing measurements respectively)}.

\resub{BAO observations measure the angular size of the baryon oscillation feature in the distribution of galaxies; the background evolution can be constrained from the evolution of the BAO feature across redshift, providing a sensitive probe of $\Omega_m$. However, such a measurement is partially degenerate with $H_0$, which can be alleviated by calibrating of the BAO standard ruler through a Big Bang Nucleosythesis (BBN) prior on the baryon density. We break the $\sigma_8-\Omega_m$ degeneracy in our analysis} by adding BAO data\editA{, primarily from the Baryon Oscillation Spectroscopic Survey (BOSS),} to our analysis we obtain
\beq \sigma_8=0.815 \pm 0.012 \hspace{0.3cm} ({\rm 3x2pt} + {\rm BAO}),\eeq also a $\sim$$1.5\%$ measurement. This value is again in good agreement with model predictions derived from $\Lambda$CDM fits to primary CMB data ($\sigma_8 = 0.8107\pm 0.0064$; the posteriors are shown in the $\sigma_8$-$\Omega_m$ plane in the right panel of Fig.\,\ref{fig:omegam_sigma8_3x2pt}).

Following the approach suggested in \cite{2021MNRAS.501.1823B} we use our CMB lensing and cross-correlation data to place constraints on the Hubble constant which arise exclusively from the measurement of the angular size of the matter-radiation equality scale imprinted in the turnover of the matter power spectrum. These constraints are independent of the sound horizon scale, knowledge of which is crucial for inferring the Hubble constant from the baryon oscillation feature either in the primary CMB or through BAO observations. We find \beq H_0=66.5^{+3.2}_{-3.7} \hun \hspace{0.3cm} ({\rm 3x2pt})\eeq from CMB lensing, cross-correlation, and galaxy auto-correlation data alone (using both ACT and \textit{Planck}). When additionally including uncalibrated supernovae from the \texttt{Pantheon+} data set \citep{2022ApJ...938..110B} to further break the degeneracy between $H_0$ and $\Omega_m$ we obtain \beq H_0=64.3^{+2.1}_{-2.4} \hun \hspace{0.3cm} ({\rm 3x2pt} + {\rm SN}).\eeq This represents a $\sim$$20\%$ improvement over the constraint presented in \cite{2024ApJ...962..113M} from the lensing auto-spectrum alone. Our results are consistent with the value of $H_0$ preferred by BAO and primary CMB data (at the $\sim$$1\sigma$ level), but in tension ($\sim$$3.6\sigma$) with local measurements of $H_0$ from \texttt{SH0ES} \citep{2024arXiv240408038B}. This is consistent with the results found using similar methods on three-dimensional galaxy clustering data \citep[see e.g.,][]{2022PhRvD.105f3503F,2022PhRvD.106f3530P,2023JCAP...04..023B}.

\editB{Posteriors for the key $\Lambda$CDM parameters ($H_0$, $\Omega_m$, $\sigma_8$, and $S_8^{\rm 3x2pt}$) probed in our analysis are shown in Fig.\,\ref{fig:3x2pt_lcdm_triangle}. We show posteriors for the 3x2pt analysis jointly using ACT and \textit{Planck} lensing data, as well as each lensing data set separately.}

Furthermore, we explore a model-agnostic parametrisation of the growth of perturbations. Using the combination of lensing auto- and cross-correlations together with BAO we find tight ($\lesssim 4\%$) constraints on $\sigma_8(z)$ in three redshift bins $z=0-1.15$, $z=1.15-2.4$, and $z>2.4$. Our reconstruction of $\sigma_8(z)$ is in good agreement with predictions within a $\Lambda$CDM model fit to the primary CMB from \textit{Planck}.

In going beyond the standard $\Lambda$CDM model we explore various extensions, including non-minimum mass neutrinos, evolving dark energy and curvature. \editB{We find constraints on the neutrino mass sum from the combination of our 3x2pt data with BAO and primary CMB data from \textit{Planck} consistent with results from analyses using only the lensing auto-spectrum}. When including both ACT and Planck lensing auto- and cross-spectra we constrain the sum of the neutrinos to 
\begin{equation}
    \begin{split}
        \sum m_\nu <& 0.124 \text{eV\ at\ 95\% confidence} \\ &({\rm 3x2pt} + {\rm CMB} + {\rm BAO}).
    \end{split}
\end{equation} 
These constraints are degraded when also considering a time varying dark energy equation of state but by additionally including uncalibrated supernovae from \texttt{Pantheon+} we still obtain a constraint of $\sum m_\nu < 0.231 \rm eV$ (95\% C.I.).

We constrain the curvature of the Universe to 
\begin{equation}
    \begin{split}
        -0.011 < \Omega_k < 0.004&\text{\ at\ 95\%\ confidence} \\ &({\rm 3x2pt} + {\rm CMB})
    \end{split}
\end{equation} 
from the primary CMB and our data alone (without BAO), \editA{about 20\%} tighter than previous results from the primary CMB and the CMB lensing auto-spectrum only \citep{2024ApJ...962..113M}. 

\editA{We make our data and likelihood publicly available enabling the community to perform further investigation into models not explored in this work (see Appendix~\ref{app:data} for details).}

\section{Data} \label{sec:data}

In this section we briefly describe the data sets we use in our analysis as well as several external likelihoods we include to break degeneracies and probe both the $\Lambda$CDM model as well as potential extensions with all available data.
%In this work we combine measurements of the large scale matter clustering from gravitational lensing of the CMB, the clustering of photometrically selected galaxies, and the cross-correlation between these probes. Furthermore, we include several external data sets to break degeneracies and probe both the $\Lambda$CDM model as well as potential extensions with all available data.

\subsection{CMB lensing power spectrum}
Throughout we adopt the CMB lensing power spectrum measurements from ACT DR6 \citep{2024ApJ...962..112Q,2024ApJ...962..113M} and \textit{Planck}~PR4 \citep{2022JCAP...09..039C}.

The ACT DR6 lensing reconstruction covers $9400\,\rm{deg}^2$ of the sky and is signal-dominated on \editB{lensing scales of $L<150$}. This reconstruction is based on CMB measurements made between 2017 and 2021 (relying only on the night-time data) at $\sim$90 and $\sim$150 GHz and uses CMB scales $600<\ell<3000$. The use of cross-correlation based estimators reduces the sensitivity to the modelling of the instrumental noise \citep{2020arXiv201102475M}, and profile hardening is employed to mitigate against extragalactic foregrounds \citep{2024ApJ...966..138M, 2020PhRvD.102f3517S,2014JCAP...03..024O}. Since the CMB lensing signal is reconstructed using quadratic estimators the power spectrum of the reconstruction is a four-point function containing several biases which need to be subtracted using simulations. The largest of these biases is the Gaussian disconnected bias, which depends on the two-point power spectrum of the observed CMB maps and is thus non-zero even in the absence of lensing. The debiasing is discussed in detail in \cite{2024ApJ...962..112Q}. Since some of the bias corrections as well as the normalisation of the lensing estimator depend weakly on cosmology we implement corrections capturing the dependency of the lensing normalisation and bias subtraction on cosmological parameters (see Appendix~\ref{app:lklh_corr} for details). The CMB lensing power spectrum from \cite{2024ApJ...962..112Q} is determined at $2.3\%$ precision, corresponding to a measurement signal-to-noise ratio of $43\sigma$.

Potential sources of systematic biases have been investigated in detail in \cite{2024ApJ...966..138M} and \cite{2024ApJ...962..112Q} and were found to be comfortably subdominant to statistical uncertainties.

The \textit{Planck} PR4 lensing analysis \citep{2022JCAP...09..039C} reconstructs lensing with CMB angular scales from $100\leq\ell\leq2048$ using the quadratic estimator. This analysis is based on the reprocessed PR4 \texttt{NPIPE} maps that incorporated around $8\%$ more data compared to the 2018 \textit{Planck} PR3 release. It also includes pipeline improvements such as optimal (anisotropic) filtering of the input CMB fields resulting in an increase of the overall signal-to-noise ratio by around $20\%$ compared to  \textit{Planck} PR3 \citep{2020A+A...641A...8P} and a detection of the lensing power spectrum  at $42\sigma$.

The $C_\ell^{\kappa \kappa}$ bandpowers for ACT DR6 and \textit{Planck} PR4 are shown in the lower panels of Fig.\,\ref{fig:bandpowers_wFit}.

\subsection{CMB lensing-galaxy cross-correlations and galaxy-galaxy auto-correlation}
\begin{figure*}
    \includegraphics[width = \linewidth, trim=0cm 0.25cm 0cm 0cm, clip]{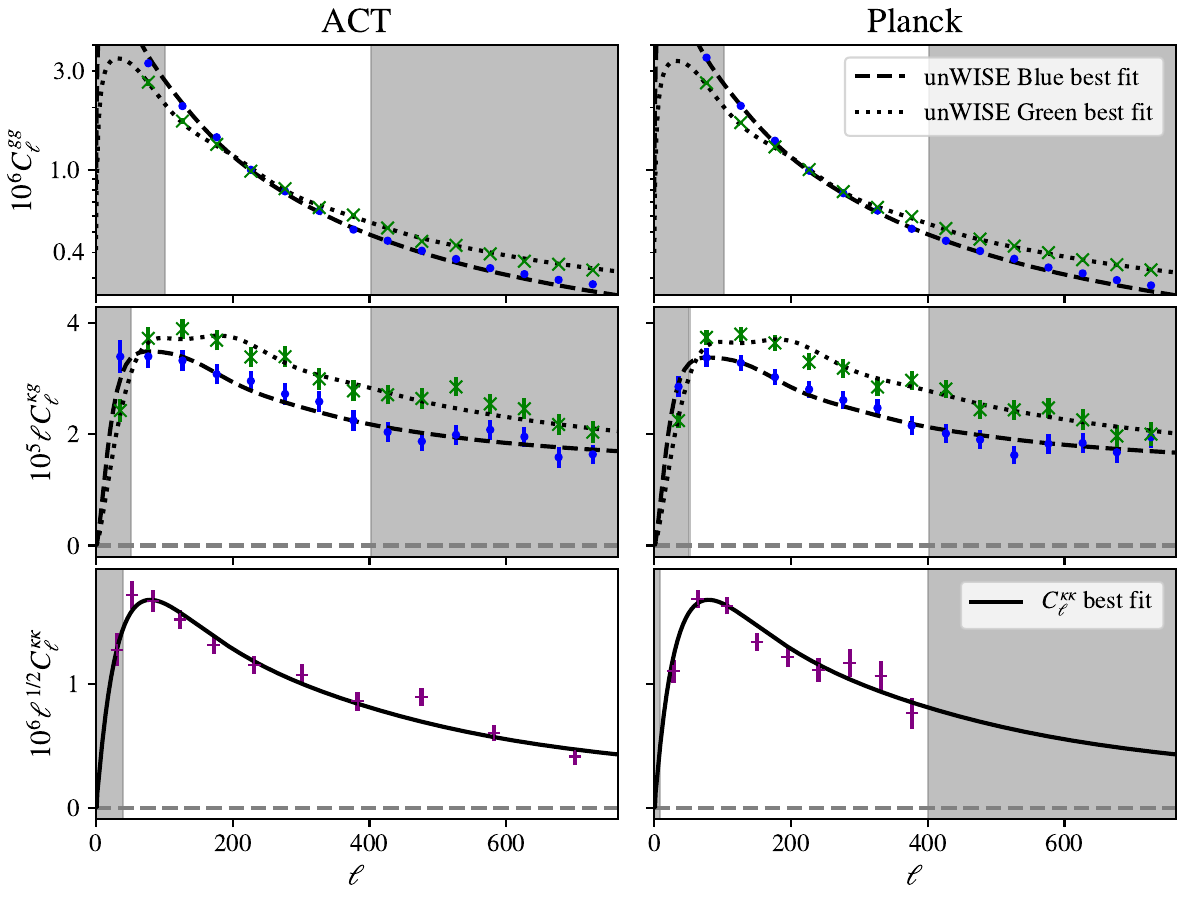}
    \caption{We show here all bandpowers which enter our joint 3x2pt analysis using ACT DR6 and \textit{Planck} PR4 CMB lensing data. From top to bottom the three panels show the galaxy-galaxy autospectra for the Blue and Green samples of unWISE galaxies measured on the ACT (left) and \textit{Planck} lensing footprints, the galaxy-CMB lensing cross-correlations, and the CMB lensing auto-correlations. We measure $C_\ell^{gg}$ independently on the ACT and \textit{Planck} areas to ensure that the galaxy selection, which can vary over the survey footprint, is consistent with the corresponding corss-correlation. The correlation between the spectra is self-consistently included in the covariance. Alongside the bandpowers we show the bestfitting spectra obtained from a joint fit of all data sets shown here and the BAO data described in Sec.\,\ref{subsec:bao_data}. We find an excellent fit to the data and estimate the probability to exceed (PTE) at $0.7$ (see Sec.\,\ref{subsubsec:goodness_of_fit}).}
    \label{fig:bandpowers_wFit}
\end{figure*}

We include measurements of the cross-correlation between CMB lensing (from ACT~DR6 and \textit{Planck}~PR4) and galaxies from the unWISE catalog, along with the auto-correlation of the unWISE galaxies \citep[from][]{2024ApJ...966..157F}. These measurements are shown in the upper two panels of Fig.\,\ref{fig:bandpowers_wFit}.

The unWISE galaxy catalogue is constructed from three band imaging by the Wide-Field Infrared Survey Explorer (WISE) survey \citep{2010AJ....140.1868W}, including four years of the post-hibernation NEOWISE phase \citep{2011ApJ...731...53M,2014ApJ...792...30M}. We employ two colour selected galaxy samples which we refer to as Blue and Green, with broad redshift distributions centred at $z \sim 0.6$ and 1.1. These samples are extensively described in \cite{2020JCAP...05..047K} and \cite{2019ApJS..240...30S}. To obtain redshift distributions for these samples of galaxies we employ the method of clustering redshifts estimated from cross-correlations with spectroscopic samples from the Sloan Digital Sky Survey (SDSS) (see \cite{2020JCAP...05..047K} and \cite{2024ApJ...966..157F} for further details).

\resub{The mode-decoupled $C_\ell^{gg}$ and $C_\ell^{\kappa g}$ for each of the two galaxy samples with the ACT and \textit{Planck} lensing reconstructions are measured using \texttt{NaMaster} \cite{2019MNRAS.484.4127A}. As discussed in \cite{2024ApJ...966..157F} these spectra are corrected for various observational effects, including galaxy survey systematics and misnormalisation of the lensing estimator. We apply a minimum scale cut of $\ell_{\rm min} = 100$ in $C_\ell^{gg}$  and $\ell_{\rm min}=50$ in $C_\ell^{\kappa g}$ motivated by our null-tests in \cite{2024ApJ...966..157F} to guard against residual contamination from large scale systematics. The small-scale cut of $\ell_{\rm max}=400$ in both $C_\ell^{\kappa g}$ and $C_\ell^{gg}$ has been verified on simulations and shown to allow for unbiased recovery of cosmological parameters.}

As with the lensing power spectrum measurements, \cite{2024ApJ...966..157F} undertook an extensive evaluation of potential systematic biases using different versions of the lensing reconstructions and analysis masks, as well as realistic simulations of the contamination due to extragalactic foregrounds. They did not find any evidence for statistically significant biases.

\subsection{External Likelihoods} \label{subsec:external_data}
We also combine our results with external data sets from observations of the primary CMB temperature and polarisation anisotropies, Baryon Acoustic Oscillations, and uncalibrated supernovae.

\subsubsection{Primary CMB} \label{subsec:primary_CMB_data}
We jointly analyse our data with measurements of the temperature and polarisation power spectra of the primary CMB as observed by the \textit{Planck} satellite. In keeping with \cite{2024ApJ...962..113M} we adopt the analysis of the \textit{Planck}~PR4 maps based on the \texttt{CamSpec} likelihood for the small scale ($\ell>30$) temperature and polarisation power spectra \citep{2022MNRAS.517.4620R}. Additionally, we include the \textit{Planck}~PR3 likelihood for the large scale temperature power spectrum \citep{2020A+A...641A...5P}. Finally, to include information from \textit{Planck}'s large-scale polarisation data that constrains the optical depth to reionisation we use the likelihood estimated in \cite{2020A+A...635A..99P} from the \texttt{SRoll2} maps.

%As an alternative to the full primary CMB likelihood we also present some constraints which use only a CMB derived prior on the parameter combination $\Omega_m h^3$. It can be shown that within $\Lambda$CDM this combination is determined by the angular size of the sound horizon and therefore largely independent from the broadband shape of the CMB power spectra \citep{2002MNRAS.337.1068P}. The combination is constraint to better than $0.3\%$ by the primary CMB data discussed above. We implement this as a Gaussian prior of $\Omega_m h^3 = (9.587 \pm 0.026) \times 10^{-2}$ \footnote{Note that this constraint is obtained from chains run with the \texttt{CamSpec} likelihood for \textit{Planck} PR4 and the \cite{2020A+A...635A..99P} likelihood for the low-$\ell$ $EE$ and $TE$ spectra, it differs slightly from the constraint reported in \cite{2020A+A...641A...6P} and the fiducial value used in \cite{2024ApJ...966..157F}.}. This is comparable to the approach taken in \cite{2024ApJ...966..157F} \citep[and previously in][]{2021JCAP...12..028K} where this parameter combination was fixed to the best fit from \textit{Planck}.

\subsubsection{Baryon Acoustic Oscillations} \label{subsec:bao_data}
Furthermore, we add observations of the BAO feature. We use a combination of BAO measurements based on the clustering of galaxies with samples spanning redshifts up to $z\simeq 1$, including 6dFGS \citep{2011MNRAS.416.3017B}, SDSS DR7 Main Galaxy Sample \citep[MGS;][]{2015MNRAS.449..835R}, BOSS DR12 LRGs \citep[Luminous Red Galaxies;][]{2017MNRAS.470.2617A}, and eBOSS DR16 LRGs \citep{2017MNRAS.470.2617A}. \editB{In contrast to earlier work on the ACT DR6 CMB lensing auto-spectrum \citep{2024ApJ...962..113M},} we additionally include the higher-redshift ELGs \citep[Emission Line Galaxies;][]{2016A+A...592A.121C}, Lyman-$\alpha$ forest \citep{2020ApJ...901..153D}, and quasar samples \citep{2021PhRvD.103h3533A} from eBOSS DR16.

\resub{As described above BAO data probes the distance-redshift relation in the late universe, providing information on the expansion rate and matter density.}

As this work was nearing completion the Year-1 BAO results from the Dark Energy Spectroscopic Instrument \citep[DESI;][]{2022AJ....164..207D} were released \citep{2024arXiv240403001D,2024arXiv240403000D,2024arXiv240403002D}. These provide improved BAO measurements tightening constraints on the cosmic expansion history. We do not reanalyse all our results with the DESI BAO, but because \cite{2024arXiv240403002D} showed that this data favours tight limits on the neutrino mass, we address our neutrino mass constraints also with the recent DESI BAO likelihood (see Sec.\,\ref{subsubsec:nulcdm}). \editB{Given the consistency and similar constraining power on $\Omega_m$ from DESI Y1 and earlier BAO data sets in the context of flat $\Lambda$CDM we do not expect the choice of BAO likelihoods to impact our constraints on structure formation.}

\subsubsection{Uncalibrated Supernovae} \label{subsubsec:sn}

Finally, we present some constraints which additionally employ `uncalibrated' measurements of the relationship between the apparent brightness of Type IA supernovae and their redshifts from the \texttt{Pantheon+} data set \citep{2022ApJ...938..110B}. Here, `uncalibrated' refers to the fact that the absolute magnitudes of the supernovae have not been calibrated, e.g., with Cepheid variables or the tip-of-the-red-giant-branch (TRGB) technique, such that only information from the relative (not absolute) distance-redshift relation is included. \resub{Supernovae observations probe the distance-redshift relation (although, in the uncalibrated case, only up to an overall amplitude) and thereby provide information on the expansion rate as a function of redshift. The addition of this data thus provide} matter density information independently of the primary CMB and BAO, which enables estimates of the Hubble constant independent of the sound horizon scale as proposed in \cite{2021MNRAS.501.1823B}. Uncalibrated supernovae also constrain potential dark energy evolution and break degeneracies between evolving dark energy and neutrino mass when considering multiple extensions to the baseline $\Lambda$CDM model jointly.

\section{3$\times$2pt Covariance} \label{sec:covmat}
As in \cite{2024ApJ...966..157F}, we use a simulation derived covariance for the CMB lensing and lensing cross-correlation measurements. \resub{Our simulation suite consists of 400 CMB lensing simulations for ACT DR6 and 480 simulations for \textit{Planck} PR4.} Ref. \cite{2024ApJ...966..157F} described how to obtain Gaussian realisations of the galaxy field that exhibit the correct correlations with lensing reconstruction simulations from \cite{2024ApJ...962..112Q} for ACT and \cite{2022JCAP...09..039C} for \textit{Planck}. They also detailed the analytic estimates for the cross-covariance between different galaxy samples and the simulations used to estimate the covariance between ACT and \textit{Planck} cross-correlations, which are obtained by running the ACT lensing reconstruction pipeline on \textit{Planck} simulations. For this work we additionally adopt the bias subtracted CMB lensing auto-power spectra measured on the same simulations to estimate the cross-covariance between the cross-correlation and auto-correlation measurements.

\resub{To generate the simulations and compute the analytic covariance we adopt the same fiducial cosmology used in the ACT lensing simulations which is based on a $\Lambda$CDM model fit to \textit{Planck}~2015 temperature and polarisation data with an updated $\tau$ prior as in~\cite{2017PhRvD..95f3525C}. To closely match the observed galaxy-galaxy and galaxy-CMB lensing spectrum we fit the bias (and other nuisance) parameters of the galaxy clustering model to the observed spectra in a manner blind to cosmology as described in \cite{2024ApJ...966..157F}.}

\editB{We find significant correlations of up to 50-60\% between the ACT lensing auto-spectrum and the cross-correlations with the ACT lensing map and slightly smaller ($\lesssim$40\%) between the respective \textit{Planck} spectra. Correlations between $C_\ell^{gg}$ and lensing auto-spectra and between cross- and auto-spectra from ACT and \textit{Planck} respectively (or vice versa) are small ($\lesssim$20\%).} \resub{ACT and \textit{Planck} largely use non-overlapping CMB scales to reconstruct the CMB lensing signal. Therefore this cross-covariance primarily arises from the signal contribution due to the partial sky overlap and a significant reduction of correlations is expected compared to the correlations between cross- and auto-spectra using the same lensing reconstructions.}

\resub{As in \cite{2024ApJ...962..112Q,2024ApJ...962..113M} and \cite{2024ApJ...966..157F} we apply a Hartlap correction to the covariance to correct for the fact that the inverse covariance estimate from a finite number of simulations is not unbiased. This correction inflates the covariance by $\sim$10\% for the analysis including only ACT or \textit{Planck} lensing reconstructions and $\sim$20\% for the joint analysis. We note that this is a conservative estimate of the Hartlap correction, since it is based on the number of CMB lensing simulations and neglects the fact that the galaxy simulations are only partially correlated with those simulations and therefore the effective number of independent simulations is larger.}

Furthermore, when analysing the low redshift data only (i.e., not in combination with primary CMB data) we propagate uncertainties in the lensing estimator normalisation due to the uncertainty in the CMB two-point functions into additional contributions to the covariance as described in \cite{2024ApJ...966..157F} and \cite{2024ApJ...962..112Q} for the cross- and auto-spectra respectively. When combining with primary CMB data, we explicitly correct for errors in the normalisation and reconstruction bias subtraction as described in Appendix~\ref{app:lklh_corr}.

\section{Cosmological analysis} \label{sec:analysis}

We obtain cosmological constraints by constructing a Gaussian likelihood
\beq
-2 \ln \mathcal{L} \propto \sum_{bb'}\begin{bmatrix}\Delta \hat{C}_b^{gg} \\ \Delta \hat{C}_b^{\kappa g} \\ \Delta \hat{C}_b^{\kappa \kappa} \end{bmatrix} \mathbb{C}^{-1}_{b b'} \begin{bmatrix}\Delta \hat{C}_{b'}^{gg} \\ \Delta \hat{C}_{b'}^{\kappa g} \\ \Delta \hat{C}_b^{\kappa \kappa}\end{bmatrix} 
\eeq
where the $\Delta \hat{C}_b^{gg}$, $\Delta \hat{C}_b^{\kappa g}$, and $\Delta \hat{C}_b^{\kappa \kappa}$ are the residuals between our observed galaxy-galaxy, galaxy-CMB lensing, and CMB lensing-CMB lensing spectra, $\hat{C}_b^{gg}$, $\hat{C}_b^{\kappa g}$, and $\hat{C}_b^{\kappa \kappa}$, and the respective binned and band window convolved theory spectra, $C_b^{gg}$, $C_b^{\kappa g}$, and $C_b^{\kappa \kappa}$. The covariance $\mathbb{C}$ has the form
\beq
\mathbb{C}_{b b'} = \begin{bmatrix}\mathbb{C}_{b b'}^{gg-gg} & \mathbb{C}_{b b'}^{gg-\kappa g} & \mathbb{C}_{b b'}^{gg-\kappa \kappa}\\ 
\left(\mathbb{C}_{b b'}^{gg-\kappa g}\right)^T & \mathbb{C}_{b b'}^{\kappa g-\kappa g} & \mathbb{C}_{b b'}^{\kappa g-\kappa \kappa} \\
\left(\mathbb{C}_{b b'}^{gg-\kappa \kappa}\right)^T & \left(\mathbb{C}_{b b'}^{\kappa g-\kappa \kappa}\right)^T & \mathbb{C}_{b b'}^{\kappa \kappa-\kappa \kappa}
\end{bmatrix}
\eeq
where $\mathbb{C}_{b b'}^{gg-gg}$, $\mathbb{C}_{b b'}^{\kappa g-\kappa g}$, and $\mathbb{C}_{b b'}^{gg-\kappa g}$ are the galaxy auto-spectrum covariance, the galaxy-CMB lensing cross-spectrum covariance, and the cross-covariance between them. $\mathbb{C}_{b b'}^{gg-\kappa \kappa}$ and $\mathbb{C}_{b b'}^{\kappa g-\kappa \kappa}$ are the cross-covariance between the galaxy-galaxy and galaxy-CMB lensing spectra on one hand and the lensing auto spectrum on the other hand. Finally, $\mathbb{C}_{b b'}^{\kappa \kappa-\kappa \kappa}$ is the CMB lensing power spectrum covariance. These are estimated from simulations as described above in Sec.\,\ref{sec:covmat}. When combining the lensing power spectrum and cross-spectrum likelihood with that for the CMB anisotropy power spectra, we ignore the covariance between the measured lensing and anisotropy spectra, as these are negligible for DR6 noise sensitivities~\citep{2013PhRvD..88f3012S, 2017PhRvD..95d3508P}. 

We infer cosmological parameters via Markov Chain Monte Carlo (MCMC) methods performing Metropolis-Hastings sampling using the \texttt{cobaya}\footnote{\url{https://github.com/CobayaSampler/cobaya}} code \citep{2021JCAP...05..057T}. We consider MCMC chains to be converged if the Gelman-Rubin statistic \citep{1992StaSc...7..457G,Gelman1998} satisfies $R-1 \leq 0.01$ for the cosmological parameters of interest.

\editA{We use a hybrid perturbation theory expansion to second order to model the galaxy auto and cross-spectra \editB{up to $k\sim0.3\hMpc$} (see \citealt{2024ApJ...966..157F} for details). We impose conservative priors on lensing magnification, shot noise, and the parameters of the bias expansion \editB{(second order and shear bias)} based on simulations. \editB{\cite{2024ApJ...966..157F} showed that these results are insensitive to these prior choices, higher order corrections contribute at most at the few percent level to the signal, and even neglecting all higher order terms leads to only minor shifts in inferred parameters ($\ll1\sigma$)}. We also marginalise over uncertainties in the redshift distribution of unWISE galaxies (see \citealt{2024ApJ...966..157F} for details). The CMB lensing power spectrum is modeled using the non-linear matter power spectrum from \texttt{HALOFIT} (see \citealt{2024ApJ...962..112Q} for details).}

\subsection{Priors}\label{subsec:priors}

In our baseline analysis we consider a spatially flat, $\Lambda$CDM universe with massive neutrinos of the minimum mass allowed in the normal hierarchy ($\sum m_\nu = 0.06$eV). When analysing only low redshift data (i.e., when not including observations of the primary CMB) our analysis is insensitive to the optical depth to reionisation, $\tau$, and we thus fix it to the mean value obtained in \cite{2020A+A...641A...6P}. For the remaining five cosmological parameters we adopt the priors from \cite{2024ApJ...962..113M} sampling the logarithm of the scalar perturbation amplitude, $\log(10^{10} A_s)$, the primordial spectral tilt, $n_s$, the physical density in baryons and cold dark matter, $\Omega_b h^2$ and $\Omega_c h^2$, and the angular size of the sound horizon at recombination, $\theta_{\rm MC}$. We place flat priors on all parameters except the baryon density and the primordial spectral tilt, which are only weakly constrained by our observations. We choose a prior motivated by BBN measurements of deuterium abundance from \cite{2020Natur.587..210M} for $\Omega_b h^2$ (see Table~\ref{tab:priors}). As pointed out in \cite{2024ApJ...962..113M} the primordial spectral tilt and the amplitude of fluctuations are somewhat degenerate given only a measurement of the projected lensing auto- or cross-spectra. We conservatively adopt a prior centred on but also about five times broader than the $n_s$-constraint obtained from \textit{Planck} measurements of the CMB anisotropy power spectra in the $\Lambda$CDM model~\citep{2020A+A...641A...6P}, and two times broader than constraints obtained there from various extensions of $\Lambda$CDM. To avoid exploring unphysical parts of parameter space we furthermore limit the range over which $H_0$ may vary to between $40$ and $100 \hun$. This limitation is relevant for some of the CMB lensing auto-spectrum-only runs we perform for comparison; all other analyses do not allow for such a wide range of $H_0$ values.

\editB{We note that these prior choices differ somewhat from those adopted in \cite{2024ApJ...966..157F}, where the baryon density and the spectral tilt were fixed to the \textit{Planck} best-fit values. Furthermore, this previous work fixed the parameter combination $\Omega_m h^3$, which within $\Lambda$CDM is closely related to $\theta_{\rm MC}$ \citep{2002MNRAS.337.1068P}, to the \textit{Planck} best-fit value.}

When combining our low redshift data sets with primary CMB observations discussed in Sec.\,\ref{subsec:primary_CMB_data} we remove the informative priors on $\Omega_b h^2$ and $n_s$ and additionally also sample the optical depth to reionsiation with a uniform prior. When exploring beyond-$\Lambda$CDM extensions we adopt the same priors as in \cite{2023PhRvD.107h3504A}. All priors are summarised in Table~\ref{tab:priors}.

\begin{table}
    \centering
    \begin{tabular}{ l c c}
    \hline
    \hline
    \multicolumn{3}{c}{\bf$\bm{\Lambda}$CDM cosmological parameters} \\ \hline
    Parameter   & Baseline Priors & comb. with CMB 2pt \\ \hline
    $\ln(10^{10} A_s)$ & $[1.61, 4.0]$ & $[1.61, 4.0]$\\
    $n_s$& $\mathcal{N}(0.96, 0.02)$& $[0.8, 1.2]$\\
    $\Omega_c h^2$& $[0.005, 0.99]$ & $[0.005, 0.99]$\\
    $\Omega_b h^2$& $\mathcal{N}(0.0223, 0.00036)$ & $[0.005, 0.1]$\\
    $100 \theta_{\rm MC}$ & $[0.5, 10]$ & $[0.5, 10]$\\
    $\tau$& fixed (0.0561) & $[0.01, 0.11]$\\ \hline \hline
    \multicolumn{3}{c}{\bf beyond $\bm{\Lambda}$CDM cosmological parameters} \\ \hline
       & fiducial value & prior \\ \hline
    \multicolumn{3}{l}{\bf$\bm{\nu\Lambda}$CDM } \\
    $\sum m_\nu$ [eV] & $0.06$ & $[0.0, 5.0]$\\
    \multicolumn{3}{l}{\bf$\bm{w}$CDM } \\
    $w$ & $-1$ & $[-3.0, 0.0]$\\
    \multicolumn{3}{l}{\bf$\bm{w_0w_a}$CDM } \\
    $w_0$ & $-1$ & $[-3.0, 0.0]$\\
    $w_a$ & $0.0$ & $[-3.0, 3.0]$\\
     &  & $w_0+w_a<0$\\
    \multicolumn{3}{l}{\bf$\bm{k\Lambda}$CDM } \\
    $\Omega_k$ & $0.0$ & $[-0.3, 0.3]$\\\hline \hline
    %\multicolumn{3}{l}{\bf{modified gravity}}\\
    %$\mu_0$ & $0.0$ & $[-1.5, 1.5]$\\
    %$\Sigma_0$ & $0.0$ & $[-1.5, 1.5]$\\
    % &  & $\mu_0 > 2\Sigma_0 + 1$\\\hline \hline
    \end{tabular}
    
    \caption{Parameters and priors used in this work. $\mathcal{N}( \mu, \sigma)$ indicates a Gaussian prior with mean $\mu$ and variance $\sigma^2$. Uniform priors are indicated by square brackets. For the priors adopted on the galaxy nuisance parameters we refer the reader to \cite{2024ApJ...966..157F}.}
    \label{tab:priors}
\end{table}

\resub{We adopt the priors on the nuisance parameters of the galaxy bias model from \cite{2024ApJ...966..157F}. This includes uniform, uninformative priors on the linear-order bias, a conservative Gaussian prior on the shot noise of 60\% around the Poisson expectation, and a 10\% Gaussian prior on the lensing magnification parameter centred on estimates obtained by computing the impact of small perturbations to the observed fluxes on the galaxy selection. The detailed procedure for setting the priors on the higher order bias parameters, which are informed by the difference between different numerical simulations is described in \cite{2024ApJ...966..157F}. Furthermore, priors on the redshift marginalisation parameters are derived from clustering redshift measurements and their uncertainties (as also described in \cite{2024ApJ...966..157F}).}

\section{Constraints on Cosmology} \label{sec:cosmo}

In the following section we present our constraints on cosmological parameters obtained from jointly analysing the ACT and \textit{Planck} lensing auto- and cross-spectra together with the auto-correlation of the unWISE galaxies and in some cases external data as discussed in Sec.\,\ref{subsec:external_data}. We begin by outlining our findings in the context of a flat $\Lambda$CDM model before discussing potential extensions beyond this model including non-minimum mass neutrinos, dark energy with equation of state $w \neq -1$, and spatial curvature.

\subsection{Constraints on flat $\Lambda\rm CDM$} \label{subsec:lcdm_cosmo}

In the context of a flat $\Lambda$CDM cosmology the weak lensing and galaxy clustering is primarily sensitive to the amplitude of matter density fluctuations in the late universe. Different probes derive their information from different scales and redshifts, providing complementary information on the evolution of matter density perturbations. As discussed in \cite{2024ApJ...966..157F} the lensing cross-correlations primarily derive their information from linear and moderately non-linear scales $0.05 \hMpc \lesssim k \lesssim 0.3 \hMpc$ and redshifts $z \simeq 0.2 - 1.6$. By contrast the lensing auto-spectrum is sensitive to a large range of redshifts between approximately 0.5 and 5 and is dominated by linear scales \citep[$k \lesssim 0.15 \hMpc$;][]{2024ApJ...962..112Q, 2024ApJ...962..113M,2006PhR...429....1L}. With the combination of these two data sets we can thus probe a wide range of redshifts from $z \simeq 0.2$ to $5$ while focusing mostly on linear and mildly non-linear scales.

These results can be compared for example to recent results from galaxy weak lensing surveys \citep[see e.g., ][]{2022PhRvD.105b3520A,2021A+A...646A.140H,2023PhRvD.108l3521S}. These are largely sensitive to smaller scales, but partially overlap with the lower end of the redshift range probed here \citep{2022MNRAS.516.5355A,2023MNRAS.525.5554P}. As described above, such surveys have consistently found a somewhat smaller amplitude of matter density fluctuations than expected within the flat $\Lambda$CDM model fit to the primary CMB data from \textit{Planck}.

In addition to constraints on matter density fluctuations, our data are also sensitive to the Universe's expansion which sets the distances to the observed structures. Features in the distribution of matter of known physical size can be used to constrain the distance to the observed structures independently of redshift estimates and therefore provide the opportunity to constrain the present day expansion rate of the Universe, $H_0$. \editC{Two such scales are imprinted onto the matter density field in the early universe:} The sound horizon scale, the distance a sound wave may travel in the early universe prior to recombination \editC{(up to $z \simeq 1100$)}, and the matter-radiation equality scale, related to the size of the horizon when the matter and radiation densities become equal \editC{($z \simeq 3500$)}. 

Our data are not directly sensitive to the former, as it manifests through oscillations in the power spectrum which are largely smoothed out due to the projection over a wide range of redshifts. However, the expansion rate can be estimated from the sound horizon feature using BAO observation from spectroscopic galaxy surveys (see e.g., those discussed in Sec.\,\ref{subsec:bao_data}). Such measurements suffer from significant degeneracies between $H_0$ and the global matter density $\Omega_m$. Our data, in which $\Omega_m$ and $H_0$ have a different degeneracy direction, can therefore serve to improve BAO derived $H_0$ estimates. 

On the other hand, our data \textit{is} directly sensitive to the latter scale which is related to the turnover of the matter power spectrum. It was pointed out in \cite{2021MNRAS.501.1823B} that CMB lensing data could therefore be used to obtain constraints on the expansion rate independent of the sound horizon scale. This scale has been the target of several modifications to the $\Lambda$CDM model which aim to resolve the tension between CMB derived $H_0$ estimates (which also depend on the sound horizon) and local measurements using the distance ladder \citep[e.g.,][]{2022ApJ...934L...7R}. This measurement equally suffers from the $H_0-\Omega_m$ degeneracy, but the direction of the degeneracy varies with redshift and with the scales probed. Combining lensing auto- and cross-spectrum information thus partially alleviates this degeneracy, allowing us to derive a constraint on $H_0$ from our CMB lensing and galaxy clustering data alone. However, the degeneracy is more effectively broken by the addition of uncalibrated supernovae data.

\subsubsection{Constraints on structure growth} \label{subsubsec:structure_growth}

Within the $\Lambda$CDM model the parameter combination best constrained by the combination of CMB lensing auto-spectrum data and our cross-correlation measurements is what we have defined as $S_8^{\rm 3x2pt}\equiv \sigma_8 (\Omega_m/0.3)^{0.4}$. With lensing data from ACT alone (ACT DR6 $C_\ell^{\kappa \kappa}$ + ACT DR6 $\times$ unWISE $C_\ell^{\kappa g}$ + unWISE $C_\ell^{gg}$) we constrain this parameter combination to $\sim$$1.8\%$,\footnote{\editA{Here and for the remainder of the section, we will label the 3x2pt analyses using the ACT DR6 or \textit{Planck} PR4 CMB lensing data sets (or their combination) as \editB{${\rm ACT}+{\rm unWISE}$ or $\textit{Planck}+{\rm unWISE}$ (${\rm ACT}+\textit{Planck}+{\rm unWISE}$) respectively. \editC{In the subsequent sections where we only consider the joint 3x2pt data set using ACT and \textit{Planck} CMB lensing data we will simply refer to these as `3x2pt'.} Meanwhile, we label the primary CMB from \textit{Planck} PR4 simply as `CMB'.}}}
\begin{equation}
    S_8^{\rm 3x2pt} = 0.819 \pm 0.015 \hspace{0.3cm}({\rm ACT}+{\rm unWISE}).
\end{equation}
This can be compared to results using $\textit{Planck}$~PR4 lensing data only 
\begin{equation}
    S_8^{\rm 3x2pt} = 0.803 \pm 0.015 \hspace{0.3cm}(\textit{Planck}+{\rm unWISE}),
\end{equation}
a $1.9\%$ constraint. The combination of ACT and \textit{Planck} data tightens these constraints by a factor of $\sim$$1.25$ to
\begin{equation}
    S_8^{\rm 3x2pt} = 0.815 \pm 0.012 \hspace{0.3cm}({\rm ACT}+\textit{Planck}+{\rm unWISE}),
\end{equation}

For comparison the best constrained parameters in the lensing-only analysis is $S_8^{\rm CMBL}\equiv \sigma_8 (\Omega_m/0.3)^{0.25}$ while the cross-correlation analysis with unWISE using $C_\ell^{\kappa g}$ and $C_\ell^{gg}$ best constrains $S_8^{\times} \equiv \sigma_8 (\Omega_m/0.3)^{0.45}$. These parameters are constrained to $2.2\%$ and $1.7\%$ from those analyses respectively (using ACT and Planck data jointly in both cases) and improvements on these parameters by moving to the 3x2pt analysis are small.

\begin{figure*}
    \centering
    \includegraphics[width=0.7\linewidth, trim=0cm 0.5cm 0cm 0cm, clip]{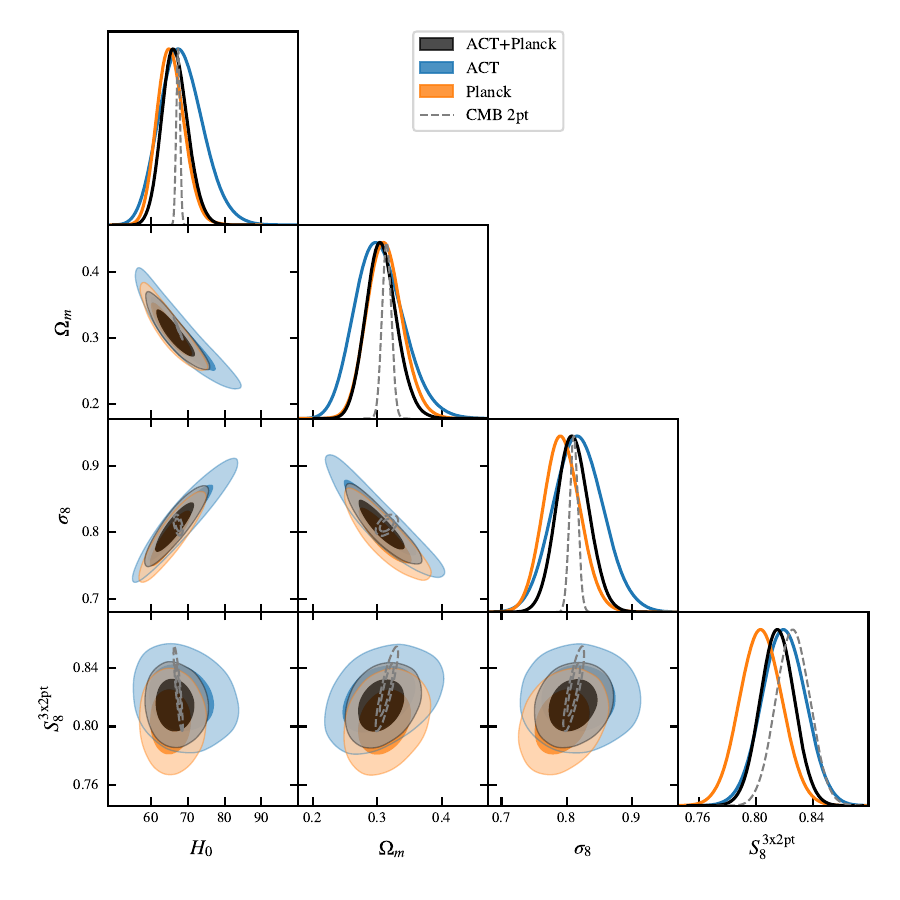}
    \caption{Posteriors on key parameters from our 3x2pt analysis without external data sets compared to parameter posteriors from a flat $\Lambda$CDM model fit to the primary CMB from \textit{Planck}. Our results on the amplitude of structure and the Hubble rate are in good agreement with the primary CMB \editB{from \textit{Planck} PR4 (shown as a dashed grey contour)} and we constrain the former \editA{independently} to the same precision.}
    \label{fig:3x2pt_lcdm_triangle}
\end{figure*}

For better comparability with galaxy weak lensing surveys, which commonly report the parameter combination $S_8\equiv\sigma_8 (\Omega_m/0.3)^{0.5}$, we also present constraints on this parameter. We find the following constraints
\begin{eqnarray}
    S_8 &=& 0.820 \pm 0.021 \hspace{0.3cm}({\rm ACT}+{\rm unWISE}),\\
    S_8 &=& 0.806 \pm 0.018 \hspace{0.3cm}(\textit{Planck}+{\rm unWISE}),\text{\ and}\\
    S_8 &=& 0.816 \pm 0.015 \hspace{0.3cm}({\rm ACT}+\textit{Planck}+{\rm unWISE}).
\end{eqnarray}
We show the constraints from our 3x2pt analysis using ACT, \textit{Planck} or ACT+\textit{Planck} comparison with primary CMB constraints from \textit{Planck} in Fig.\,\ref{fig:3x2pt_lcdm_triangle}.

To directly constrain $\sigma_8$ we need to break the degeneracy with $\Omega_m$. We have two ways of doing this: either by adding BAO data which provide constraints primarily in the $\Omega_m-H_0$ parameter space or by adding uncalibrated supernovae which constrain $\Omega_m$. With BAO we find
\begin{eqnarray}
    \sigma_8 &=& 0.815 \pm 0.013\\ \nonumber &&({\rm ACT}+{\rm unWISE}+ {\rm BAO}),\\
    \sigma_8 &=& 0.806 \pm 0.013\\ \nonumber &&(\textit{Planck}+{\rm unWISE} + {\rm BAO}),\text{\ and}\\
    \sigma_8 &=& 0.815 \pm 0.012\\ \nonumber &&({\rm ACT}+\textit{Planck}+{\rm unWISE} + {\rm BAO})
\end{eqnarray}
compared to 
\begin{eqnarray}
    \sigma_8 &=& 0.796 \pm 0.020\\ \nonumber &&({\rm ACT} +{\rm unWISE} + {\rm SN}),\\
    \sigma_8 &=& 0.780 \pm 0.019\\ \nonumber &&(\textit{Planck} +{\rm unWISE} + {\rm SN}), \text{\ and}\\
    \sigma_8 &=& 0.794 \pm 0.016\\ \nonumber &&({\rm ACT}+\textit{Planck} +{\rm unWISE} + {\rm SN})
\end{eqnarray}
with supernovae data.

Our constraints in combination with BAO are primarily limited by lack of knowledge of $n_s$. As noted above we adopted a conservative prior of $\sigma(n_s)=0.02$ which is about five times wider than the constraint from \textit{Planck} \citep[$n_s=0.9649 \pm 0.0044$;][]{2020A+A...641A...6P}. We investigate instead adopting a more aggressive prior of $\sigma(n_s)=0.01$ corresponding approximately to the largest $1\sigma$-posterior found for various beyond-$\Lambda$CDM extensions examined by \textit{Planck} \citep{2020A+A...641A...6P}. This tightens the constraints on $\sigma_8$
\begin{eqnarray}
    \sigma_8 &=& 0.815 \pm 0.012\\ \nonumber &&({\rm ACT} +{\rm unWISE} + {\rm BAO}),\\
    \sigma_8 &=& 0.805 \pm 0.012\\ \nonumber &&(\textit{Planck} +{\rm unWISE} + {\rm BAO}),\text{\ and}\\
    \sigma_8 &=& 0.814 \pm 0.010\\ \nonumber &&({\rm ACT}+\textit{Planck} +{\rm unWISE} + {\rm BAO}).
\end{eqnarray}
%Gravitational lensing observations that are sensitive to the total 
%matter distribution
As we can see the constraints from ACT and \textit{Planck} data alone, while affected by the $n_s$ prior, are less sensitive to this choice than the combination of both, which is increasingly limited by our conservative prior and improved by $\sim$$20\%$ when tightening the prior on $n_s$.

Given the agreement with primary CMB predictions reported in \cite{2024ApJ...962..112Q,2024ApJ...962..113M} and \cite{2024ApJ...966..157F} we do not expect strong disagreement with \textit{Planck} constraints from the primary CMB. Indeed, our lensing constraints are  consistent with the primary CMB data sets discussed in Sec.\,\ref{subsec:primary_CMB_data}. From the primary CMB we have $S_8^{\rm 3x2pt} = 0.826 \pm 0.012$ ($S_8=0.830 \pm 0.014$, $\sigma_8=0.8107 \pm 0.0064$) about $0.6\sigma$ ($0.7\sigma$, $0.3\sigma$ or $1\sigma$ comparing to the analyses with BAO or SN respectively) away from the joint result reported above. 

We also combine our data with the primary CMB to further break degeneracies. We find even tighter constraints on $\sigma_8$ of
\begin{eqnarray} 
    \sigma_8 &=& 0.8124 \pm 0.0048\\ \nonumber &&({\rm ACT} +{\rm unWISE} + {\rm CMB}),\\
    \sigma_8 &=& 0.8105 \pm 0.0047\\ \nonumber &&(\textit{Planck} +{\rm unWISE} + {\rm CMB}),\text{\ and}\\
    \sigma_8 &=& 0.8127 \pm 0.0044\\ \nonumber &&({\rm ACT}+\textit{Planck} +{\rm unWISE} + {\rm CMB}).\label{eq:sigma8_wCMB2pt}
\end{eqnarray}
\editA{This constraint represents a $\sim$$30\%$ improvement over the constraint from \textit{Planck} primary CMB data alone, demonstrating that with precise knowledge of the scalar spectral index, the matter density and other cosmological parameters our data provides a powerful probe of matter density fluctuations.} \editB{We also note, however, that this result only represents a marginal, $\sim$2.5\% improvement over the combination of \textit{Planck} primary CMB data with the ACT and \textit{Planck} lensing auto-spectra. This is not entirely unexpected given that within the $\Lambda$CDM model low redshift structure formation is constrained tightly by the lensing auto-spectrum alone.}
%\SF{Perhaps worth pointing out that this is $\sim 40\%$ better than Planck 2018! So 3x2pt helps a lot when you know $n_s$ rather well...}
\subsubsection{Constraints on Hubble expansion}\label{subsubsec:hubble}

As discussed above our data can be used in two ways to shed light on the expansion rate of the Universe, $H_0$. Firstly, when combining our data with BAO to break the degeneracy between $H_0$ and $\Omega_m$ we find
\begin{eqnarray}
    H_0 &=& 67.53\pm 0.80 \hun\\ \nonumber &&({\rm ACT} +{\rm unWISE} + {\rm BAO} + {\rm BBN}),\\
    H_0 &=& 67.24\pm 0.78 \hun\\ \nonumber &&(\textit{Planck} +{\rm unWISE}+ {\rm BAO} + {\rm BBN}),\text{\ and}\\
    H_0 &=& 67.35\pm 0.78 \hun\\ \nonumber &&({\rm ACT}+\textit{Planck} +{\rm unWISE} + {\rm BAO} + {\rm BBN}).
\end{eqnarray}
\editC{We note that these constraints depend sensitively on the BBN prior on the baryon density discussed in Sec.\,\ref{subsec:priors} which is crucial in determining the sound horizon size.} These results are consistent with results from the primary CMB ($H_0=67.32 \pm 0.51 \hun$  from the \textit{Planck} primary CMB), approximately $20\%$ tighter than results from BOSS BAO alone ($H_0=67.33\pm0.98\hun$ when including the BBN prior used in this work as well; \citealt{2021PhRvD.103h3533A})\editA{, and comparable to the constraint from the newer DESI BAO measurements ($H_0=68.53\pm0.80\hun$ when calibrating the sound horizon ruler with BBN information; \citealt{2024arXiv240403002D}).}

\begin{figure}
    \centering
    \includegraphics[width=\linewidth]{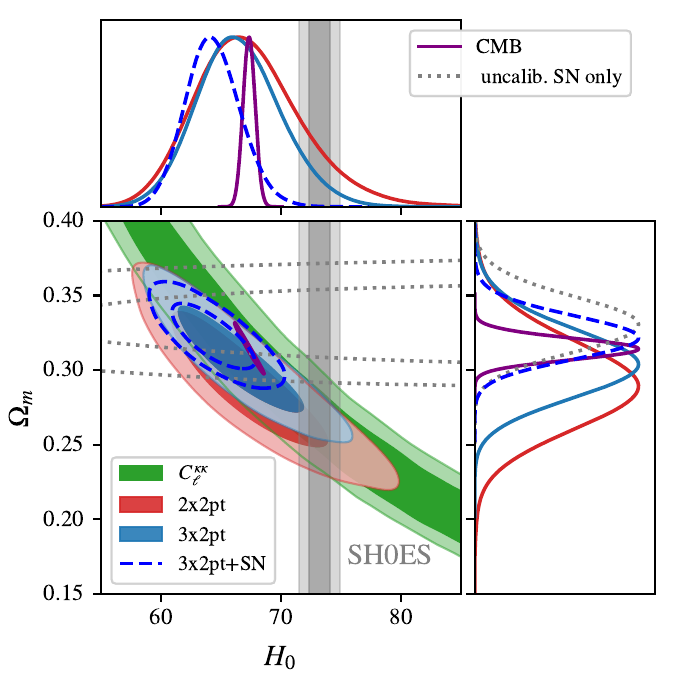}
    \caption{\editB{With 3x2pt data we are able to constrain the Hubble constant $H_0$ to approximately 5\% based on the matter-radiation equality scale without external data due to slightly different degeneracy directions in the cross- and auto-correlations. By adding uncalibrated supernovae data we further break the $H_0$-$\Omega_m$ degeneracy and improve constraints to about 3.5\% showing a $\sim$3.6$\sigma$ tension with $H_0$ measurements from the Cepheid-calibrated distance ladder from \texttt{SH0ES} \citep[shown as a grey band;][]{2024arXiv240408038B}.}}
    \label{fig:H0_OmegaM_3x2pt}
\end{figure}

\editA{Secondly, we can use the fact that, as discussed above, our data are not directly sensitive to the sound horizon to place independent constraints on the Hubble rate through measurements of the matter-radiation equality scale. %Such sound horizon independent constraints are of interest because modifications to the physical size of the sound horizon have been a common target for attempts to alleviate the Hubble tension.
} From the 3x2pt data alone we find
\begin{eqnarray}
    H_0 &=& 68.4^{+5.2}_{-6.5} \hun\\ \nonumber &&({\rm ACT} +{\rm unWISE}),\\
    H_0 &=& 65.4^{+3.3}_{-4.0} \hun\\ \nonumber &&(\textit{Planck} +{\rm unWISE}),\text{\ and}\\
    H_0 &=& 66.5^{+3.2}_{-3.7} \hun\\ \nonumber &&({\rm ACT}+\textit{Planck} +{\rm unWISE}).
\end{eqnarray}
When combining our data with uncalibrated supernovae to further break the $H_0-\Omega_m$ degeneracy we obtain
\begin{eqnarray}
    H_0 &=& 64.8^{+2.6}_{-3.0} \hun\\ \nonumber &&({\rm ACT}+{\rm unWISE} + {\rm SN}),\\
    H_0 &=& 63.5^{+2.2}_{-2.4} \hun\\ \nonumber &&(\textit{Planck}+{\rm unWISE} + {\rm SN}),\text{\ and}\\
    H_0 &=& 64.3^{+2.1}_{-2.4} \hun\\ \nonumber &&({\rm ACT}+\textit{Planck}+{\rm unWISE} + {\rm SN}).
\end{eqnarray}
\editB{The posteriors on $H_0$ and degeneracies with $\Omega_m$ are shown in Fig.\,\ref{fig:H0_OmegaM_3x2pt}.}

\subsubsection{Comparison of $\Lambda$CDM constraints with external analyses}

\begin{figure}
    \centering
    \includegraphics[width=\linewidth]{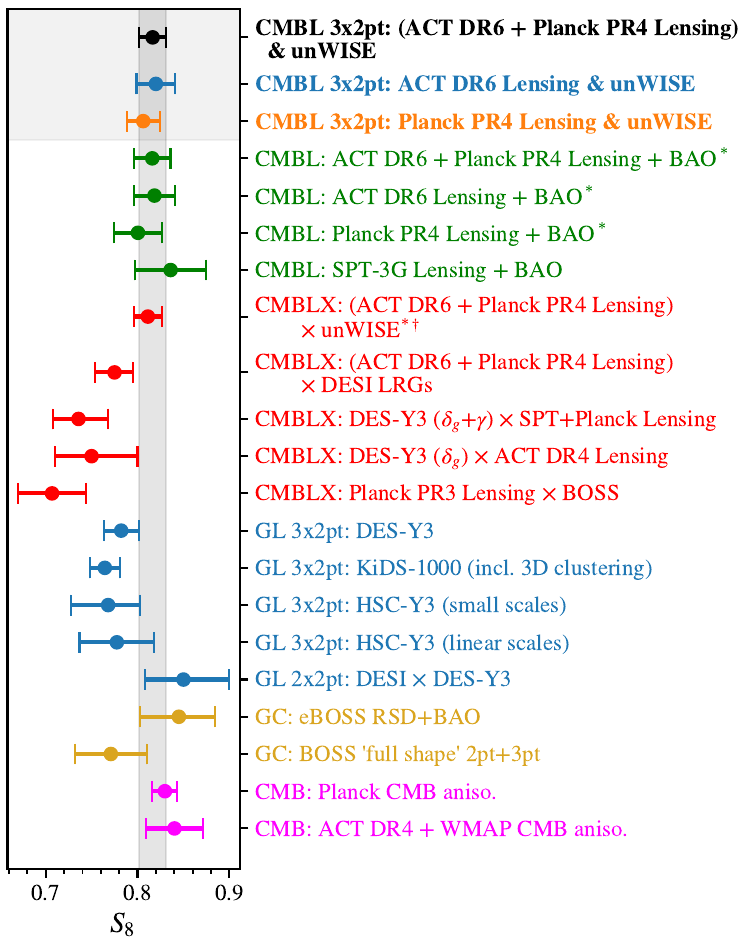}
    \caption{Here we compare an extensive set of measurements of $S_8$ \editC{to our results shown in the shaded box at the top of the figure}. We include measurements from CMB lensing in combination with BAO (in green; though only the \editB{SPT-3G} result is independent from our 3x2pt analysis), from cross-correlations between CMB lensing and galaxy surveys (in red; again some of these measurements are included in our 3x2pt analysis), from 3x2pt analyses of galaxy weak lensing (in blue), from the three dimensional clustering for spectroscopic galaxy surveys (in gold), and predictions from the primary CMB (in magenta). Our results are in good agreement with CMB lensing analyses, as well as with the predictions from the primary CMB. While we favour a slightly larger amplitude of matter density fluctuations than \resub{most} galaxy weak lensing analyses and many of the CMB lensing cross-correlations we compare to, we do not find any strongly significant discrepancies. This also holds for the $S_8$ inferred from the three dimensional clustering of spectroscopic galaxies.\\
    $^*$ Denotes data sets included in our 3x2pt analysis.\\
    $^\dagger$ Note that these are not identical to the results presented in \cite{2024ApJ...966..157F} as we have re-analysed them with the priors used in this work. The difference in mean $S_8$ is, however, less than $0.1\sigma$ with identical uncertainties.}
    \label{fig:comp_S8_whisker}
\end{figure}

\begin{figure}
    \centering
    \includegraphics[width=\linewidth]{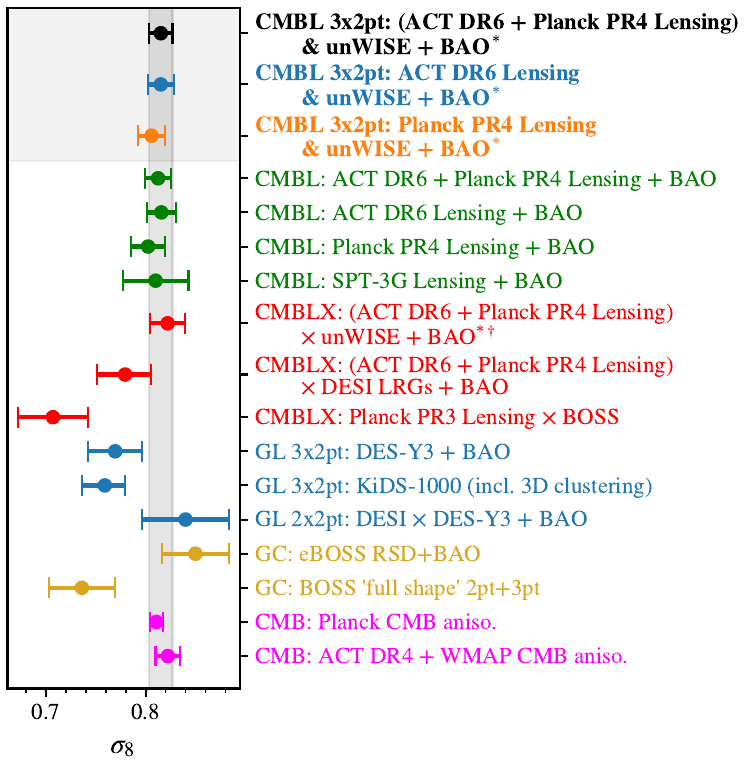}
    \caption{Similarly to Fig.\,\ref{fig:comp_S8_whisker} we also compare a set of measurements of $\sigma_8$. We again include measurements from CMB lensing in combination with BAO (in green; though only the \editB{SPT-3G} result is independent from our 3x2pt analysis), from cross-correlations between CMB lensing and galaxy surveys (+BAO) (in red; again some of these measurements are included in our 3x2pt analysis), from 3x2pt analyses of galaxy weak lensing also in combination with BAO (in blue), from the three dimensional clustering for spectroscopic galaxy surveys (in gold), and predictions from the primary CMB (in magenta). As in the case of $S_8$ we find good agreement with CMB lensing analyses, as well as with the predictions from the primary CMB. The agreement with other large scale structure tracers is broadly similar to the $S_8$ parameter; we favour a slightly larger value of $\sigma_8$ but only at moderate significance ($\sim$$1-2.5\sigma$).\\
    $^*$ Denotes data sets included in our 3x2pt analysis.\\
    $^\dagger$ Note that these are not identical to the results presented in \cite{2024ApJ...966..157F} as we have re-analysed them with the priors and BAO data sets used in this work. In $\sigma_8$ the shift is slightly larger than in $S_8$ ($0.5\sigma$) and and we find a slightly larger uncertainty (by about $20\%$).}
    \label{fig:comp_sigma8_whisker}
\end{figure}

As discussed in Sec.\,\ref{sec:key_results} our constraints on the amplitude of matter density fluctuations are in excellent agreement with model predictions from a flat $\Lambda$CDM model (with minimum neutrino mass, $\sum m_\nu = 0.06$eV) fit to the primary CMB from \textit{Planck} (see Sec.\,\ref{subsec:primary_CMB_data} for a detailed description of the data sets used). In Figs.\,\ref{fig:comp_S8_whisker}~and~\ref{fig:comp_sigma8_whisker} we show these comparisons graphically and additionally show a comparison with independent primary CMB observations from the combination of \textit{WMAP} and ACT \citep{2020JCAP...12..047A} (shown in magenta in Figs.\,\ref{fig:comp_S8_whisker}~and~\ref{fig:comp_sigma8_whisker}). 

We also provide comparisons to a wider set of other large scale structure probes. We include results from CMB lensing auto-spectrum analyses, galaxy weak lensing surveys (DES, KiDS and HSC), other cross-correlations with CMB lensing from ACT, SPT, and \textit{Planck}, and from the three dimensional clustering of galaxies from BOSS and eBOSS. We subsequently briefly introduce the datasets we employ:
\begin{itemize}
    \item \textbf{CMBL:} These are constraints from the analysis of the auto-spectrum of CMB lensing reconstructions (shown in green in Figs.\,\ref{fig:comp_S8_whisker}~and~\ref{fig:comp_sigma8_whisker}). These results are mostly sensitive to linear scales at $z=1-2$ \editA{but with a broad tail to higher redshifts}. CMB lensing primarily constrains the parameter combination $\sigma_8 \Omega_m^{0.25}$. In addition to the two CMB lensing auto-spectra which also enter our analysis, \textit{Planck} PR4 \citep{2022JCAP...09..039C}, ACT DR6 \citep{2024ApJ...962..112Q,2024ApJ...962..113M}, we also compare to the independent analysis from \editB{SPT-3G} \citep{2023PhRvD.108l2005P}. To make fair comparisons we combine the CMB lensing measurements with BAO information which breaks the degeneracy between $\sigma_8$ and $\Omega_m$.

    \item \textbf{CMBLX}: We also compare our results with cross-correlations between CMB lensing and galaxy surveys (using only the galaxy-CMB lensing cross-correlation and the galaxy-galaxy auto-correlation; 2x2pt). These are shown in red in Figs.\,\ref{fig:comp_S8_whisker}~and~\ref{fig:comp_sigma8_whisker}. We use results from \cite{Kim2024} and \cite{Sailer2024} which analysed the cross-correlation between DESI LRG targets and the ACT DR6 lensing reconstruction. Ref. \cite{Sailer2024} also present a reanalysis of the DESI LRGs' cross-correlation with \textit{Planck} PR4 updating the results from \cite{2022JCAP...02..007W}. 
    %\editA{Furthermore, we include the analysis from \cite{Qu2024} which uses the same lensing reconstructions but the full DESI Legacy Survey (DESI LS) dataset. This work includes additional DESI objects (e.g., ELGs in addition to the LRGs), but does not draw on the spectroscopic calibration of the redshift distributions.} 
    We also include work cross-correlating DES-Y3 galaxy shear ($\gamma$) and galaxy densities ($\delta_g$) with a joint SPT and \textit{Planck} lensing reconstruction \editC{based on a much smaller survey footprint than considered here} \citep{2023PhRvD.107b3530C}. The joint lensing reconstruction from SPT-SZ and \textit{Planck} is presented in \cite{2023PhRvD.107b3529O}. A cross-correlation between DES-Y3 clustering ($\delta_g$) and ACT DR4 lensing \citep{2024JCAP...01..033M} based on the lensing reconstruction from \cite{2021MNRAS.500.2250D} is also included. While all aforementioned analyses use photometric galaxy samples in their cross-correlations, the final cross-correlation study included in our comparisons, \cite{2022JCAP...07..041C}, jointly models the three dimensional clustering of the spectroscopic BOSS galaxies and their cross-correlation with \textit{Planck}.

    \item \textbf{GL:} From galaxy weak lensing surveys we include constraints from `3x2pt' analyses, combining measurements of galaxy clustering, galaxy shear, and their cross-correlation (shown in blue in Figs.\,\ref{fig:comp_S8_whisker}~and~\ref{fig:comp_sigma8_whisker}). For DES we adopt the results obtained in \cite{2022PhRvD.105b3520A} when fixing the neutrino mass. For KiDS we show results presented in \cite{2021A+A...646A.140H}. We note that in contrast to the DES analysis the KiDS results are obtained from the combination of projected galaxy shear and galaxy-galaxy lensing measurements with a measurement of the three dimensional clustering of galaxies in the spectroscopic BOSS and 2dFLenS surveys. Therefore, the KiDS analysis already contains the BAO information. For HSC we adopt a set of results based on measurements of the galaxy shear, galaxy-galaxy lensing, and galaxy clustering from \cite{2023PhRvD.108l3520M}. We show results for an analysis using a linear bias model \citep{2023PhRvD.108l3521S} and from an analysis using a halo model based emulator which includes non-linear scales \citep{2023PhRvD.108l3517M}.

    \resub{In addition to these `3x2pt' analyses we also include one `2x2pt' analysis which excludes the galaxy shear auto-correlation and uses galaxy positions from DESI and galaxy shear measurements from DES. This analysis, presented in \cite{2024PhRvD.110j3518C}, demonstrates the first self-consistent treatment of galaxy-galaxy lensing within Lagrangian perturbation theory and employs a substantially more flexible model for intrinsic alignments (IA) than usually applied in these type of analyses.}

    \item \textbf{GC}: Finally, we compare our results with constraints from the three dimensional clustering of galaxies as measured by BOSS and eBOSS (shown in gold in Figs.\,\ref{fig:comp_S8_whisker}~and~\ref{fig:comp_sigma8_whisker}). We include the analysis of BAO and Redshift Space Distortions (RSD) from \cite{2021PhRvD.103h3533A}. Furthermore, we include an independent analysis based on the effective theory of large scale structure (EFTofLSS) that fits the `full shape' of the power spectrum and bispectrum measured in redshift-space \citep{2023PhRvD.107h3515I}. Similar, previous analyses found compatible results \citep[see e.g.,][]{2020JCAP...05..005D,2020JCAP...05..042I,2022JCAP...02..008C}.
\end{itemize}

\begin{figure}
    \centering
    \includegraphics[width=\linewidth]{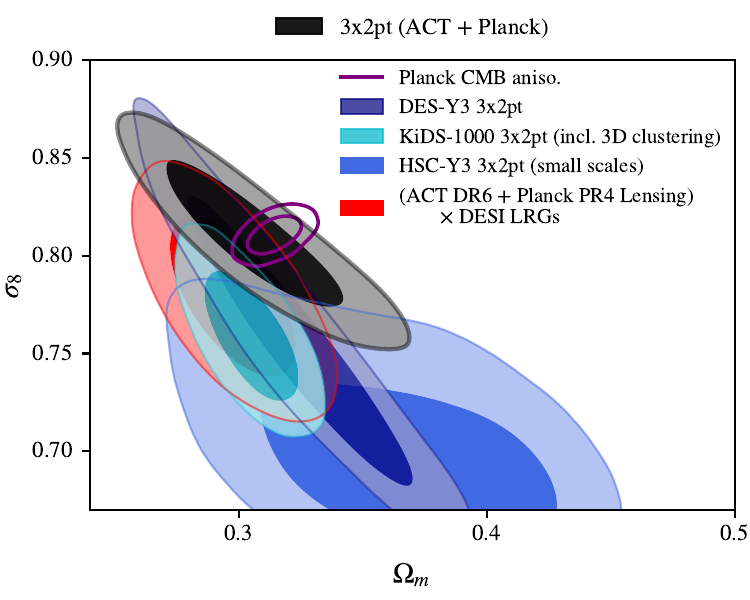}
    \caption{Within the $\sigma_8$-$\Omega_m$ plane one can see that our results show no significant tension with any one galaxy weak lensing survey. The 3x2pt contour in black shows significant overlap with the contours from DES, KiDS and HSC (blue tones). There is also overlap with the posterior from the cross-correlation between ACT DR6 and \text{Planck} PR4 CMB lensing reconstructions and DESI LRG targets (in red).}
    \label{fig:sigma8_omegam_comp}
\end{figure}

Comparisons for the $S_8$ and $\sigma_8$ parameters are shown in Figs.\,\ref{fig:comp_S8_whisker} and \ref{fig:comp_sigma8_whisker}. We also show comparisons in the $\sigma_8$-$\Omega_m$ plane for a few selected external data sets in Fig.\,\ref{fig:sigma8_omegam_comp}. Generally we find good agreement with the CMB lensing results, although we note only the estimate from \editB{SPT-3G} is not already included in our 3x2pt analysis. When comparing to the four different galaxy weak lensing 3x2pt analyses we find that these generally favour a lower value of $S_8$ at moderate significance between $\sim$$1\sigma$ and $2.3\sigma$. \resub{However, the 2x2pt galaxy lensing analysis presented in \cite{2024PhRvD.110j3518C} which applies a more general treatment of IA finds a larger values of $S_8$, consistent with the value preferred by \textit{Planck} and our results.} Meanwhile, a lower amplitude is also found in other CMB lensing cross-correlations we compare our results to. From the DESI LRG targets we find a $\sim$$1.6\sigma$ lower value of $S_8$ using CMB lensing data from ACT DR6 and \textit{Planck} PR4 ($0.9\sigma$ and $1.9\sigma$ for ACT DR6 and \textit{Planck} PR4 alone respectively). 
%\editA{A similar cross-correlation analysis using the full DESI LS galaxy sample also finds $S_8$ about $1.5\sigma$ lower than our results.} 
Using DES galaxies and cosmic shear together with the SPT+\textit{Planck} lensing reconstruction also yields a $\sim$$2.4\sigma$ lower value for $S_8$, while the discrepancy with the cross-correlation between DES and ACT DR4 is slightly less significant ($1.4\sigma$). The results from the cross-correlation of spectroscopic galaxies from BOSS with \textit{Planck} PR3 lensing also yields a value of $S_8$ that is $2.7\sigma$ lower than our inference. Meanwhile the disagreement with galaxy clustering is less significant ($0.6-1.0\sigma$).

Where available we also consider results that directly constrain $\sigma_8$ either by combining with BAO or because they include the BAO information as part of a three dimensional clustering analysis and compare these to our analysis with BAO. We broadly find similar levels of agreement/discrepancy as for $S_8$ except in the case of the `full-shape' galaxy clustering analysis which finds a value of $\sigma_8$ about $2.3 \sigma$ lower than our results.

We conclude that while our results are in good agreement with primary CMB and CMB lensing results, there are some moderate discrepancies with other large scale structure tracers from galaxy weak lensing surveys and (mostly lower redshift) CMB lensing cross-correlations. However, as we can see in Fig.\,\ref{fig:sigma8_omegam_comp}, the posteriors for several of these data sets have significant overlap with our results in the $\sigma_8-\Omega_m$ plane. The projection onto $S_8$ slightly exaggerates the level of disagreement. \editB{At current levels of precision the various data sets are in broad agreement and we are unable to conclusively rule out statistical fluctuations as the source of the observed discrepancies.}

Furthermore, we also compare our constraints on the Hubble constant to a range of external measurements. This includes model dependent measurements based on the apparent size of the sound horizon such as those from the primary CMB \citep{2020A+A...641A...6P,2020JCAP...12..047A} and BAO \citep{2021PhRvD.103h3533A}, as well as measurements based on the matter-radiation equality scale from \cite{2024ApJ...962..113M} and \cite{2022PhRvD.106f3530P}. Finally, we compare to several local measurements of $H_0$, including the TDCOSMO strong-lensing time-delay measurement with marginalisation over lens profiles \citep{2020A+A...643A.165B}, an alternative TDCOSMO measurement with different lens-mass assumptions \citep{2020A+A...643A.165B}, the Cepheid-calibrated \texttt{SH0ES} supernovae measurement \citep{2024arXiv240408038B}, the TRGB-calibrated supernovae measurement \citep{2019ApJ...882...34F}\footnote{\editC{We note that a more recent TRGB-calibrated measurement using data from the James Web Space Telescope finds similar values of $H_0$ but with slightly larger uncertainties due to a smaller sample size \citep{2024arXiv240806153F}.}}, \editC{and recent results employing calibration based on observation of asymptotic giant branch stars in the J-band \citep[JAGB;][]{2024arXiv240803474L}.}

We find good agreement with other equality scale based measurements. The constraint from the 3x2pt data alone is in excellent agreement with sound horizon based measurements while the measurements including the supernovae data set yield values about $1.3\sigma$ lower than the value inferred from the CMB, but in no statistically significant tension. At the same time the combination of 3x2pt data and supernovae is in $\sim$$3.6\sigma$ tension with local measurements of $H_0$ from the Cepheid-calibrated supernovae (\texttt{SH0ES}). \editB{We find no significant tension with TRGB- or JAGB-calibrated supernovae measurements ($\sim$$1.7\sigma$ and $\sim$$0.8\sigma$ respectively); the change is driven in part by larger uncertainties but also due to a lower central value of $H_0$ inferred with these methods.}

\begin{figure}
    \centering
    \includegraphics[width=\linewidth]{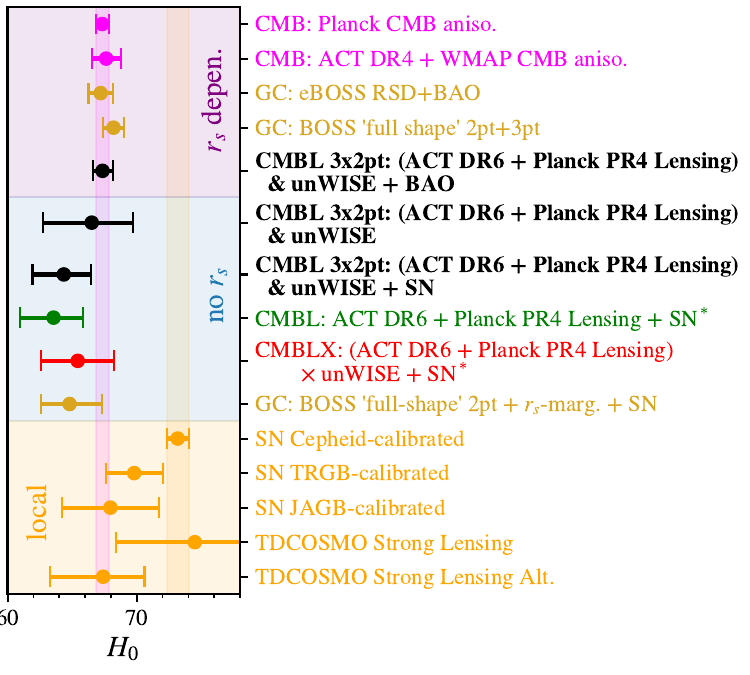}
    \caption{A comparison with various estimates of the Hubble constant. We include measurements dependent on the sound horizon (on light purple background) from the primary CMB (in magenta), the three dimensional clustering of galaxies (in gold), and our own work in combination with BAO (first black data point). On a light blue background we contrast these measurements with several sound horizon independent measurements from this work (in black), from the analysis of the CMB lensing power spectrum (in green; note that these are included in the 3x2pt results from this work), from the CMB lensing cross-correlation analysis (in red; this measurement is also part of our 3x2pt analysis), and from the clustering of BOSS galaxies with explicit marginalisation over the sound horizon scale. Finally, on a light orange background we show several results from local measurements of $H_0$ using Cepheid-, TRGB-\editB{, or JAGB-}calibrated supernovae, and two strong-lensing time-delay measurements using different lensing mass profiles.\\
    $^*$ Denotes data sets included in our 3x2pt analysis.}
    \label{fig:comp_H0_whisker}
\end{figure}

\subsubsection{\resub{Goodness of fit in the context of $\Lambda$CDM}}\label{subsubsec:goodness_of_fit}

\resub{We now turn to quantifying the goodness of fit provided by our model described above and a $\Lambda$CDM cosmology to the observed spectra. For simplicity we restrict ourselves for this purpose to our fiducial 3x2pt analysis using both ACT and \textit{Planck} CMB lensing data with and without the addition of the degeneracy-breaking BAO data. In total our data set consists of 71 bandpowers (6 for each $C_\ell^{gg}$ measured on the ACT and \textit{Planck} lensing footprints, 7 for each $C_\ell^{\kappa g}$, and 10 and 9 bandpowers for the ACT and \textit{Planck} $C_\ell^{\kappa \kappa}$ respectively). In addition to that we employ a total of 15 BAO measurements.}

\resub{The contribution of the model parameters to the effective degrees of freedom is less straightforward to estimate. Our model has a total of 41 free parameters, but only for seven of these we employ flat, uninformative priors. This includes the amplitude of scalar perturbations, $\ln(10^{10} A_s)$, the dark matter density parameter, $\Omega_c h^2$, the angular size of the sound horizon at recombination, $\theta_{\rm MC}$, as well as the linear order galaxy bias parameters. For the remaining parameters we adopt informative priors as described in Sec.\ref{subsec:priors}. Since these parameters are to varying degree informed by both the prior and the data it is not appropriate to include them as free parameters (and correspondingly reduce the number of degrees of freedom) even when the $\chi^2$-contribution from the prior is included in the total $\chi^2$. To estimate the appropriate effective number of prior degrees of freedom we fit a $\chi^2$-distribution to the observed prior $\chi^2$-values in our analysis chains. We find that in the analysis with (without) BAO data the priors contribute 28.7 (29.0) effective degrees of freedom. So that the effective number of free parameters should be taken to be 12.3 (12.0).}

\resub{We find a minimum total $\chi^2$ of 66.8 (55.0) when including (not including) the BAO data and therefore obtain a probability to exceed (PTE) of 0.70 (0.63). Hence the bestfitting model provides an excellent fit to the data.}

\resub{Finally, we can compute the Bayesian equivalent of the best-fit PTE defined as the probability that a random realisation of the data, ($\tilde{\bm {d}}$) drawn from the posterior has a larger average $\chi^2$ than the observed realisation ($\bm{d}$) \citep{2014bda..book.....G} \footnote{\resub{Such a measure may be defined for any other test statistic, but we adopt $\chi^2$ for simplicity and computational ease.}}. Since our likelihood is Gaussian it can  be shown that, given the model parameters, $\bm{\theta}$, the ${\rm PTE} \equiv {\rm Prob}\left(\chi^2(\bm{\theta}, \tilde{\bm{d}}) \geq \chi^2(\bm{\theta}, \bm{d})\vert \bm{d}\right)$ can be computed as \citep[see e.g.,][]{Sailer2024}}
\beq
{\rm PTE} = \int \dd \bm{\theta}  \left[ 1 - \frac{1}{\Gamma(N_d/2)} \gamma\left(\frac{N_d}{2}, \frac{\chi^2(\bm{\theta}, \bm{d})}{2} \right) \right] P(\bm{\theta}\vert \bm{d})\nonumber \\
\eeq
\resub{where $N_d$ is the number of data degrees of freedom and $P(\bm{\theta} \vert \bm{d})$ is the posterior on the model parameters given the observed data. We find ${\rm PTE}= 0.68$ ($0.62$) when including (not including) the BAO data and similarly conclude that the data is well described by a $\Lambda$CDM model.}

\subsection{Reconstructing the growth of perturbations}

\begin{figure}
    \centering
    \includegraphics[width=\linewidth]{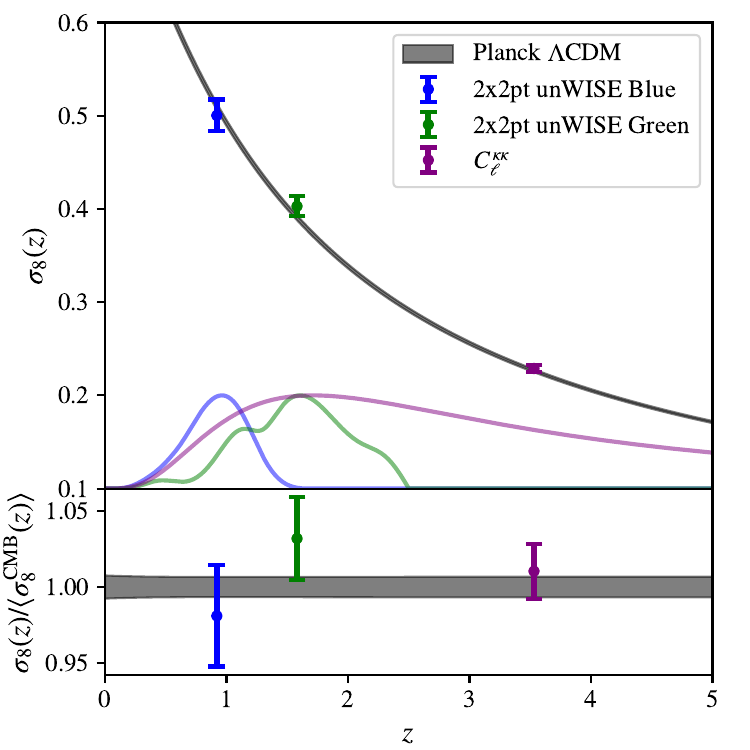}
    \caption{\editB{Our constraints on the formation of structure with redshift, $\sigma_8(z)$, are in excellent agreement with predictions from a $\Lambda$CDM model fit to the primary CMB from \textit{Planck} (shown as a grey band). We show here the constraints from the CMB lensing cross-correlations with both unWISE samples, Blue and Green, alongside the constraint from the lensing auto-correlation (in purple). In each case we show the joint constraint using ACT and \textit{Planck} CMB lensing data. The faint kernels shown at the bottom of the top panel indicate the redshift sensitivity of the three samples given by $\dd \log {\rm SNR}/ \dd z$. The ${\rm SNR}$ computation includes approximate marginalisation over the linear order galaxy bias and the shot noise \editC{as described in the body of the paper.}}}
    \label{fig:sigma8_z_recon_approx}
\end{figure}

\editB{In addition to measuring the growth of structure within the $\Lambda$CDM model through constraints on a single parameter, $S_8^{\rm 3x2pt}$ or $\sigma_8$, we are able to derive information on the evolution of structures over time by leveraging the different redshift sensitivity of the cross-correlations with the two unWISE samples and the CMB lensing auto-correlation. In Fig.\,\ref{fig:sigma8_z_recon_approx} we show the constraints from the two cross-correlations and the CMB lensing auto-correlation, each analysed jointly with BAO. The redshift kernels shown at the bottom of the top panel give an indication of the redshift sensitivity of the samples given by the fractional contribution to the signal-to-noise, $\dd \log {\rm SNR} / \dd z$. The computation of $\dd \log {\rm SNR} / \dd z$ includes an approximate marginalisation over galaxy nuisance parameters, achieved by linearising the model for $C_\ell^{gg}$ and $C_\ell^{\kappa g}$ in small fluctuations around the best-fit linear bias and shot noise and propagating the uncertainty in these parameters to the covariance matrix.}

\editB{We adopt the median of the redshift sensitivity kernel to represent the effective redshift of each of the three measurements and compute $\sigma_8(z)$ as $\sigma_8(z=0)D(z)$, where $D(z)$ is the linear growth function which is primarily dependent on $\Omega_m$. These results are also summarised in Table~\ref{tab:approx_growth_z_constraints}. We find excellent agreement with the growth of structures predicted by the $\Lambda$CDM model fit to the primary CMB from \textit{Planck} (grey band in Fig.\,\ref{fig:sigma8_z_recon_approx}).}

\begin{table*}
    \centering
    \begin{tabular}{l c| C C C C}
    \hline \hline
            & & \sigma_8(z=0) & \Omega_m & z_{\rm Med} & \sigma_8(z_{\rm Med})\\\hline
         unWISE Blue $\times$ ACT + Planck CMB lensing & $C_\ell^{g g} 
 + C_\ell^{\kappa g}$ & 0.797\pm 0.024 & 0.317\pm 0.014 & 0.9 & 0.500\pm 0.017\\
         unWISE Green $\times$ ACT + Planck CMB lensing &$C_\ell^{g g} 
 + C_\ell^{\kappa g}$ & 0.824\pm 0.019 & 0.291\pm 0.010 & 1.6 & 0.403\pm 0.011\\
         ACT + Planck CMB lensing & $C_\ell^{\kappa \kappa}$ & 0.812\pm 0.013 & 0.303\pm 0.011 & 3.5 & 0.229\pm 0.004\\\hline \hline
    \end{tabular}
    \caption{\editB{We report $\sigma_8(z)$ at the median redshifts of the three samples used in this work, the cross-correlations between CMB lensing and the unWISE Blue and Green samples and the CMB lensing auto-correlation. The median redshift is computed by estimating the fractional contribution to the signal from a given redshift, $\dd \log {\rm SNR}/ \dd z$. To obtain a faithful representation of the origin of the cosmological information content of the measurements we include approximate marginalisation over the galaxy nuisance parameters in the $\dd \log {\rm SNR}/ \dd z$ computation. As can be seen in Fig.\,\ref{fig:sigma8_z_recon_approx} the constraints from all three samples arise from a broad range of redshifts and have significant overlap.}}
    \label{tab:approx_growth_z_constraints}
\end{table*}

\editB{However, as can be easily seen from Fig.\,\ref{fig:sigma8_z_recon_approx}, the three samples have significant redshift overlap. In particular, while the median redshift of the measurement from the lensing auto-spectrum is $z_{\rm Med}\simeq 3.5$ it receives significant contributions from lower redshifts where we also have information from the cross-correlation measurements. To optimally combine the available information we explore a reconstruction of the growth of (linear) perturbations with redshift through a parametric form of $\sigma_8(z)$ which we constrain jointly with all three samples taking into account their overlap. With this method we are able to use the cross-correlation measurements to constrain the low redshift contribution to the lensing auto-spectrum and extract information on the integrated growth of structure at high redshifts, \emph{above the two galaxy samples} ($z\gtrsim 2.4$).}
%In addition to the physically motivated extensions to the $\Lambda$CDM model we discussed in the preceding sections we also investigate a model agnostic parametrisation of the growth of (linear) perturbations with redshift. 
Due to the broad redshift kernels of our data sets we cannot constrain the growth of perturbations with arbitrary resolution. Instead we adopt the following simple parametrisation similar to \cite{2023PhRvD.107h3504A}: We rescale the linear power spectrum as follows
\begin{equation}
    \begin{split}
       P_{\rm lin}^{\rm new}(k, z) =& P_{\rm lin}^{\rm input}(k, z) A(z)\\ =& P_{\rm lin}^{\rm input}(k, z) \begin{cases}
        A_0 & 0\leq z < z_1\\
        A_1 & z_1\leq z < z_2\\
        A_2 & z_2\leq z
    \end{cases}. 
    \end{split}
\end{equation}
Where $P_{\rm lin.}^{\rm input}(k, z)$ is the linear matter power spectrum computed by \texttt{CAMB} at a given cosmology and $A_0, A_1$, and $A_2$ are free parameters. The redshift bins, $z_1=1.15$ and $z_2=2.4$, are motivated by the redshift origin of the signal for the two cross-correlation measurements and are chosen to separate the two samples as optimally as possible. $A_0$ is primarily constrained by the Blue sample of unWISE galaxies, while the Green sample is primarily sensitive to $A_1$. The lensing auto-correlation receives contributions from a wide range of redshifts, but in combination with the two cross-correlation measurements allows us to constrain the amplitude at high redshift, $A_2$.

We marginalise over $A_i$ with uniform priors in the range $0$ to $2$. The parameters of interest are then $\sigma_8 \sqrt{A_i}$. Since our model depends on the non-linear \texttt{HMCode} matter power spectrum (see \cite{2024ApJ...962..112Q} and \cite{2024ApJ...966..157F} for detailed descriptions of the models used to fit the lensing auto- and cross-correlations respectively) we take 
\begin{equation}
\begin{split}
    P_{\rm non-lin.}^{\rm new}(k, z) =& P_{\rm lin.}^{\rm new}(k, z) + P_{\rm non-lin.}^{\rm input}(k, z) - P_{\rm lin.}^{\rm input}(k, z)\\ =& P_{\rm lin.}^{\rm input}(k, z) \left[A(z) - 1\right] + P_{\rm non-lin.}^{\rm input}(k, z).
\end{split}
\end{equation}

The non-linear contributions to the \texttt{HMCode} matter power spectrum have a non-trivial dependence on the amplitude of linear fluctuations. In our parametrisation $\sigma_8$ on its own (in contrast to its product with $\sqrt{A_i}$) is only constrained by the size of the non-linear terms. This approach effectively allows us to marginalise over the size of these non-linear contributions\footnote{\resub{We note that the non-linear contributions constrain $\sigma_8$ to $0.856\pm 0.055$ for the 3x2pt analysis with BAO; consistent with our fiducial analysis but with a more than 4.5 times larger uncertainty.}}.

\begin{figure*}
    \centering \includegraphics[width=0.7\linewidth,trim=0cm 0.6cm 0cm 0cm, clip]{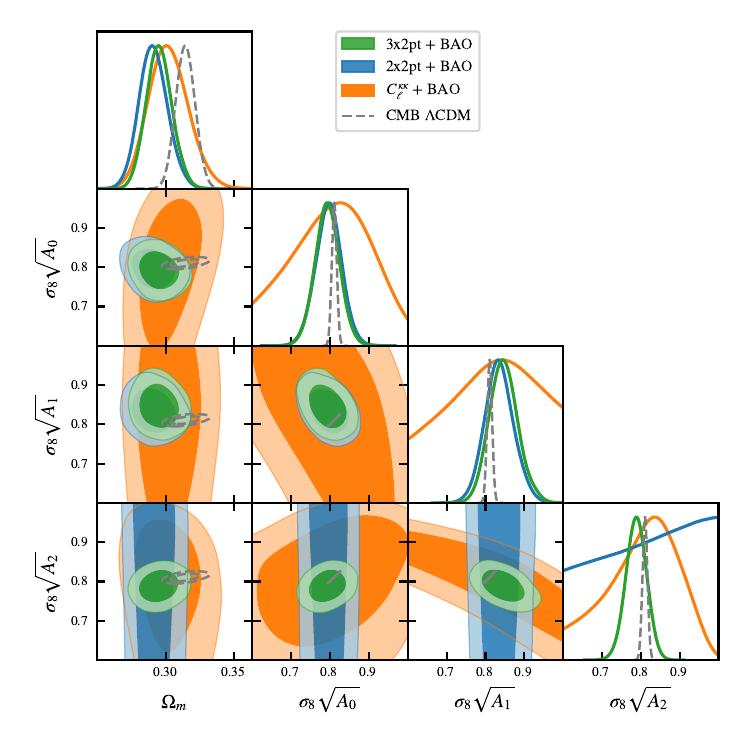}
    \caption{By rescaling the linear matter power spectrum in three redshift bins $0\leq z<1.15$, $1.15\leq z<2.4$, and $2.4\leq z$ we obtain constraints on the growth of perturbations with redshift independent of any beyond-$\Lambda$CDM model. We show here the combination of $\sigma_8$ and the square root of the rescaling parameter in the three redshift bins along with the constraints on $\Omega_m$ for our 3x2pt analysis, the 2x2pt analysis and the lensing auto-spectrum only. In each case we combine with BAO and we compare to the $\sigma_8$ predicted from the primary CMB. The amplitude of fluctuations in each of the bins is strongly degenerate when using only $C_\ell^{\kappa \kappa}$. By adding the lower redshift cross-correlation measurements we break these degeneracies and constrain $\sigma_8 \sqrt{A_i}$ to \editB{better than} 4\% in each of the bins. See also Fig.\,\ref{fig:sigma8_z_recon} for a reconstruction of $\sigma_8(z)$ from these constraints.}
    \label{fig:triangle_free_growth}
\end{figure*}

\begin{figure*}
    \centering
    \includegraphics[width=\linewidth]{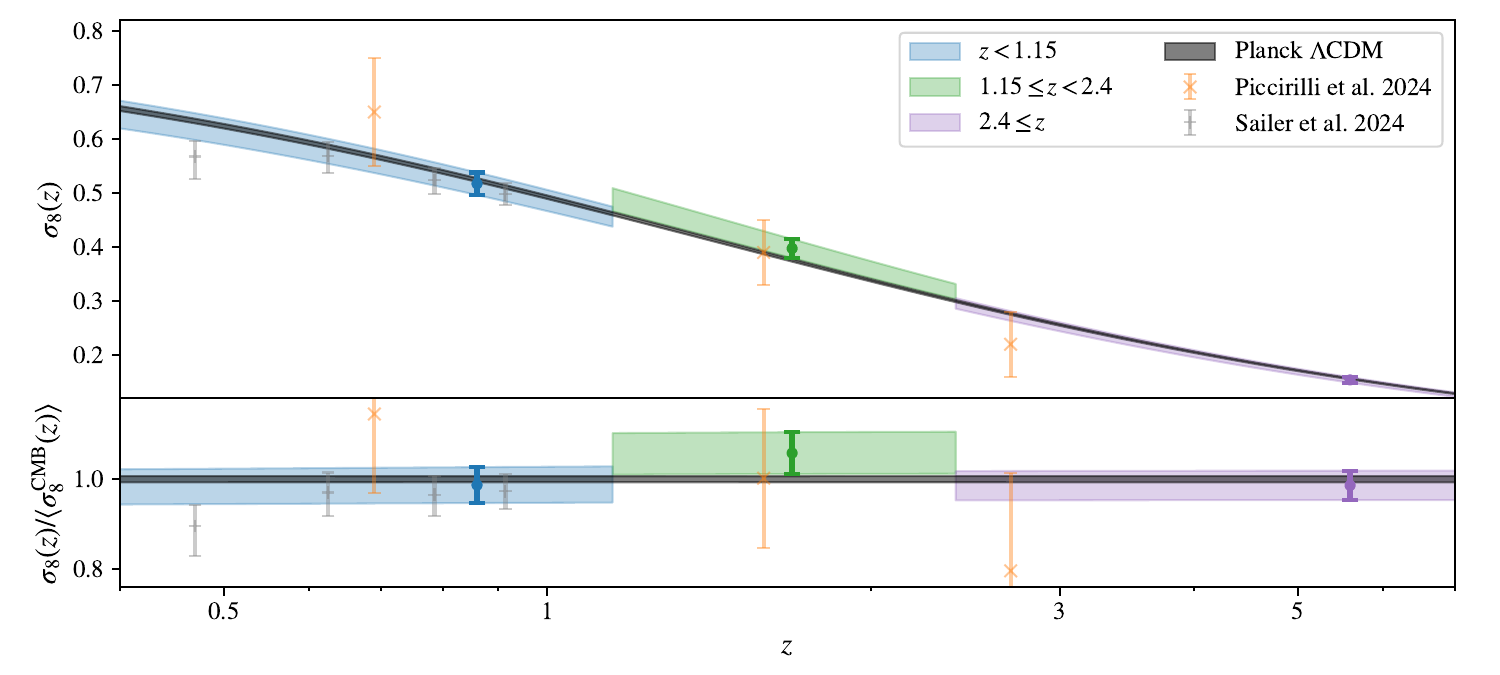}
    \caption{Using our freely scaled (linear) perturbations we can reconstruct $\sigma_8(z)$ across a wide range of redshifts. We show here the $1\sigma$ uncertainty bands on this reconstruction using the 3x2pt data with BAO and, for comparison, $\sigma_8(z)$ as predicted from a fit to the primary CMB. Note that we only constrain an overall amplitude in each of the three bins, but not the shape of the $\sigma_8(z)$ within the bin. \editB{Within each bin we highlight our constraints at the median signal redshift (blue, green and purple data points) which can be compared to other reconstructions of the growth of structure with redshift for example from the cross-correlation of CMB lensing with DESI LRGs \citep{Sailer2024} or the \textit{Quaia} quasar catalog \citep{2024JCAP...06..012P}.}}
    \label{fig:sigma8_z_recon}
\end{figure*}

Fig.\,\ref{fig:triangle_free_growth} shows the results of this analysis in the form of the posteriors on $\sigma_8 \sqrt{A_i}$ for the 3x2pt, 2x2pt and lensing-only data sets. In each case we combine with BAO data \editB{(described in Sec.\,\ref{subsec:bao_data})} to break the degeneracy between $\sigma_8$ and $\Omega_m$. We find good consistency with the amplitude of fluctuations predicted by a $\Lambda$CDM fit to the primary CMB in each of the three bins. 

We can also use these results to reconstruct $\sigma_8(z)$ which we show in Fig.\,\ref{fig:sigma8_z_recon}. \editB{We show the 1$\sigma$ confidence intervals from our inference chains on the combination $\sqrt{A_i} \sigma_8(z=0) D(z)$ within the relevant bins. \editC{We do not separately constraint the shape of $D(z)$ within each bin, but rather assume a single $D(z)$ across all bins with its shape determined primarily by $\Omega_m$.} Within each bin we compute the median redshift of the joint signal as we did above for the individual measurements. The parameter constraints are summarised in Table~\ref{tab:free_growth_constraints}.}

\editB{The representative constraints at the median signal redshift can be compared to results from the literature for example from the cross-correlation of CMB lensing with DESI LRGs \citep{Sailer2024} or the \textit{Quaia} quasar catalog \citep{2024JCAP...06..012P}. Our results are broadly consistent with these external reconstructions across the entire redshift range. At low redshifts ($z\lesssim1$) we achieve similar precision ($\sim$4\%) to constraints presented in \cite{Sailer2024} but cannot match the redshift resolution of the DESI samples. In the redshift range $1 \lesssim z  \lesssim 2.5$ our results represent some of tightest constraint on the amplitude of matter density fluctuations; tighter than results in a similar redshift range from \cite{2024JCAP...06..012P} by a factor of about 4. The main power of our method, however, is the ability to place constraints on structure formation at higher redshifts, typically inaccessible with cross-correlation analysis. The median signal redshift within our highest redshift bin is $z_{\rm Med} = 5.6$, significantly higher than the highest redshift constraint available from cross-correlation with \textit{Quaia} ($z=2.7$). In this high redshift range we obtain a $\sim$4\% constraint on the amplitude of matter density fluctuations, consistent with predictions based on a $\Lambda$CDM fit to the primary CMB from \textit{Planck}. } 

As one can see in Fig.\,\ref{fig:triangle_free_growth} some residual correlations remain between the redshift bins. We find that bin 1 and 2 are about 37\% correlated while bin 2 and 3 are 45\% correlated. The correlation between bin 1 and 3 is about 24\%.

When using only the lensing auto- and cross-correlations as well as the unWISE auto-correlation, but without BAO, $\sigma_8 \sqrt{A_i}$ is degenerate with $\Omega_m$, as expected. In each of the three bins we therefore determine the best constrained $S_8$-equivalent combination for our 3x2pt data alone. Using all lensing data from ACT and \textit{Planck} we find
\begin{eqnarray}
    \sigma_8 \sqrt{A_0} (\Omega_m/0.3)^{0.66} &=& 0.784\pm 0.035,\\
    \sigma_8 \sqrt{A_1} (\Omega_m/0.3)^{0.49} &=& 0.839\pm 0.042,\text{\ and}\\
    \sigma_8 \sqrt{A_2} (\Omega_m/0.3)^{0.39} &=& 0.783\pm 0.033.
\end{eqnarray}
We can see that, as expected, the lower redshift results are more degenerate with $\Omega_m$.

\begin{table}
    \centering
    \begin{tabular}{C| C C C}
    \hline \hline
    & \sigma_8(z=0) \sqrt{A_i} & z_{\rm Med} & \sigma_8(z_{\rm Med}) \sqrt{A_i}\\ \hline
    z<1.15 & 0.793\pm 0.032 & 0.9 & 0.517\pm 0.021\\
    1.15 \leq z < 2.4 & 0.847^{+0.034}_{-0.038} & 1.7 & 0.397\pm0.018\\
    2.4 \leq z & 0.788\pm 0.027 & 5.6 & 0.154\pm 0.005\\\hline \hline
    \end{tabular}
    \caption{\editB{We summarise here constraints on the amplitude of linear matter density fluctuations as a function of redshift from our 3x2pt analysis using ACT and \textit{Planck} CMB lensing data in combination with BAO. Alongside the inferred constraint on $\sigma_8 \sqrt{A_i}$ at $z=0$. We also provide a constraint at an representative, median redshift for each bin. To compute this we again consider the nuisance-marginalised fractional $SNR$ contributions as a function of redshift.}}
    \label{tab:free_growth_constraints}
\end{table}

\subsection{Constraints on beyond $\Lambda\rm CDM$ cosmologies}

We explore several extensions beyond the standard flat $\Lambda$CDM cosmology with minimum mass neutrinos. Generally we combine our data for this purpose with primary CMB observations. We also add BAO and supernovae data in some cases. Unless otherwise noted the constraints reported in the following sections consider the 3x2pt data set using both ACT and \textit{Planck} lensing data. We summarise the constraints on the beyond $\Lambda$CDM parameters in Table~\ref{tab:beyond_lcdm_summary}.

\begin{table*}
    \centering
    \begin{tabular*}{\linewidth}{@{\extracolsep{\fill}} c | C | C C C | C C}
        \hline \hline
         \multicolumn{2}{c|}{~}& \specialcell[c]{\text{3x2pt}\\\text{+CMB}} & \specialcell[c]{\text{3x2pt}\\\text{+CMB}\\\text{+BAO}} & \specialcell[c]{\text{3x2pt}\\\text{+CMB}\\\text{+BAO}\\\text{+SN}} & \specialcell[c]{\text{Ext. only:}\\\text{CMB}} & \specialcell[c]{\text{Ext. only:}\\\text{+CMB}\\\text{+BAO}\\\text{+SN}}\\ \hline
         \hspace{-0.6cm}\specialcell[c]{$\Lambda$CDM\\+ $\sum m_\nu$} \hspace{0.6cm}& \sum m_\nu \text{\ [eV]}& <0.219 & <0.123 & <0.137 & <0.373 & <0.163 \\\hline
         $w$CDM & w & -1.53^{+0.20}_{-0.31} & -1.026^{+0.051}_{-0.043} & -0.982 \pm 0.025 & -1.48^{+0.19}_{-0.42}& -0.979\pm 0.026\\ \hline
         \multirow{3}{*}{$w_0w_a$CDM} & w_0 & -1.24^{+0.50}_{-0.59} & -0.56 \pm 0.24 & -0.881 \pm 0.059 & -1.23^{+0.49}_{-0.60}& -0.888\pm 0.062\\ 
         ~& w_a & - & -1.17^{+0.70}_{-0.61}&-0.40^{+0.25}_{-0.21} & -&-0.40^{+0.27}_{-0.24}\\
         ~& w_p & -1.57^{+0.28}_{-0.36} & -1.011^{+0.056}_{-0.051} & -0.980 \pm 0.026 & -1.52^{+0.28}_{-0.45} & -0.976 \pm 0.028\\\hline
         k$\Lambda$CDM & 10^2\Omega_k & -0.30^{+0.41}_{-0.36} & 0.00 \pm 0.17& -& -2.07^{+1.3}_{-1.0}& 0.00 \pm 0.16\\\hline \hline
    \end{tabular*}
    \caption{Summary of beyond $\Lambda$CDM parameter constraints. When providing two sided constraints we provide the 68\% confidence intervals. For the neutrino mass we report a 95\% upper limit as in the body of the paper. \editA{We did not perform the joint analysis of our 3x2pt data, CMB, BAO and SN for the k$\Lambda$CDM cosmology as the analysis including non-vanishing curvature is computationally expensive and SN data are not expected to improve constraints on $\Omega_k$.} \editB{We compare our constraints to those from external data sets only: the CMB alone and a combination of CMB, BAO, and SN.}}
    \label{tab:beyond_lcdm_summary}
\end{table*}

\subsubsection{$\nu\Lambda\rm CDM$}\label{subsubsec:nulcdm}

Observations of neutrino flavour oscillations require that neutrinos have non-vanishing mass. Since any mechanism to give neutrinos mass requires physics beyond the standard model, this has significant implications for our understanding of fundamental physics. However, oscillation experiments only constrain the difference between the squared masses of the three mass eigenstates, $\Delta_{12} m^2 \equiv m_1^2 - m_2^2$ and $\vert \Delta_{32} m^2 \vert$ where $\vert \Delta_{32} m^2 \vert \gg \Delta_{12} m^2$, but not the absolute mass scale. Current constraints dictate that the sum of the neutrino masses $\sum m_\nu$ is \textit{at least} 0.058eV in what is known as the normal hierarchy ($\Delta_{32} m^2>0$; two of the masses are significantly smaller than the third) or \textit{at least} about 0.1eV in the inverted hierarchy ($\Delta_{32} m^2<0$; two of the masses are significantly larger than the third).

Direct detection experiments like \texttt{KATRIN} \citep[see recent results in][]{2019PhRvL.123v1802A,2022NatPhys...2..160A,2024arXiv240613516A} have placed upper limits on the effective electron anti-neutrino mass, $m_{\nu_e}<0.45$eV at 90\% confidence, from tritium beta decay observations. However, the most stringent limits on the sum of the neutrino masses are derived from cosmological observations. After neutrinos become non-relativistic they cluster similarly to CDM on scales above the neutrino free streaming length, but on smaller scales they suppress the growth of perturbations due to their large velocity dispersion. Meanwhile the neutrino energy density also contributes to the Universe's background expansion. The scale dependence of the suppression effect is only poorly constrained by current generation data like ours, but we are sensitive to the overall suppression of the formation of structure.

For this purpose we combine our lensing auto- and cross-spectrum information which probes the matter power spectrum at late times, after neutrinos have become non-relativistic, with primary CMB data which provides information on the early time matter power spectrum before suppression due to massive neutrinos can manifest. We furthermore also add BAO information to probe the background expansion and break degeneracies, for example with the total matter density. It should be noted that the amplitude of the primordial power spectrum extracted from the small scale CMB power spectrum is completely degenerate with the optical depth to reionisation, $\tau$. It is thus important to include the large scale CMB polarisation data from the \texttt{SRoll2} analysis discussed in Sec.\,\ref{subsec:primary_CMB_data}.

We extend the flat $\Lambda$CDM model by a single free parameter, $\sum m_\nu$, corresponding to the sum of the neutrino masses\footnote{As in \cite{2024ApJ...962..113M} we follow \cite{2006PhR...429..307L} and \cite{2018JCAP...04..017D} and consider a degenerate combination of three equally massive neutrinos.}. Using ACT and \textit{Planck} lensing and cross-correlation data we find an upper limit on the sum of the neutrino masses of
\begin{equation}
    \begin{split}
        \sum m_\nu <& 0.124 \text{eV\ at\ 95\% confidence} \\ &({\rm 3x2pt} + {\rm CMB} + {\rm BAO}).
    \end{split}
\end{equation}

\begin{figure}
    \centering
    \includegraphics[width=\linewidth, trim=0cm 0.6cm 0cm 0cm, clip]{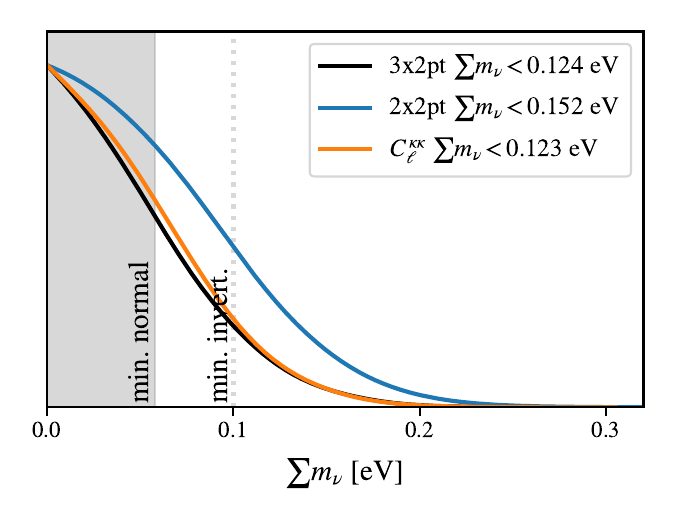}
    \caption{Using the combination of our low redshift lensing data with primary CMB anisotropies and BAO we constrain the sum of the neutrino masses to $\sum m_\nu < 0.124$ eV\ at\ 95\% confidence using lensing data from ACT and \textit{Planck}. For comparison we show here also the constraints arising from using only the lensing auto-spectrum or only the cross-correlations.}
    \label{fig:mnu_constraints}
\end{figure}

This represents a small improvement over the constraint reported in \cite{2024ApJ...962..113M} ($\sum m_\nu < 0.13 \text{eV;\ 95\% C.I.}$) based on the lensing power spectrum only. However, the difference is driven by the additional, higher redshift, BAO likelihoods used in this work which favour a lower value of $\Omega_m$. When reanalysing the lensing auto-spectrum  with these likelihoods we find $\sum m_\nu < 0.123 \text{eV;\ 95\% C.I.}$ (see also Fig.\,\ref{fig:mnu_constraints}). 

\resub{This is expected since CMB lensing probes a combination of the matter density fluctuations and the total matter density which enters through the lensing kernel. As previously discussed, the best constrained parameter combination (for the CMB lensing auto-spectrum, as well as for our 3x2pt analysis) is of the form $\sigma_8 \Omega_m^{\gamma}$, so that a lower value for $\Omega_m$ increases $\sigma_8$, compatible with less neutrino-induced structure suppression and thus a lower neutrino mass}. 

The lack of improvement from the 3x2pt analysis over auto-spectrum only constraints is also not unexpected given that we find a slightly smaller amplitude of matter density fluctuations in the combined analysis of lensing auto- and cross-correlation data than from the auto-correlation alone, compatible with slightly more suppression due to massive neutrinos and therefore a larger neutrino mass. On idealised mock observations we show that, assuming minimum mass normal hierarchy neutrinos, the 3x2pt analysis leads to a \editA{marginally} ($\sim$5\%) tighter upper limit on the neutrino mass sum than the $C_\ell^{\kappa \kappa}$-only analysis.

When including supernovae observations in addition to BAO and primary CMB, the data favours a slightly larger matter density. This results in a lower inference for $\sigma_8$ (see also results in Sec.\,\ref{subsubsec:structure_growth}) and compatibility with larger neutrino-induced power suppression. Hence the posterior on $\sum m_\nu$ is shifted to larger values which leads to a small degradation in the one sided constraint to
\begin{equation}
    \begin{split}
        \sum m_\nu <& 0.137 \text{eV\ at\ 95\% c.l.} \\ &({\rm 3x2pt} + {\rm CMB} + {\rm BAO} + {\rm SN}).
    \end{split}
\end{equation}

When replacing our baseline BAO likelihoods from 6dFGS, SDSS, BOSS, and eBOSS with the new DESI BAO likelihood we find significantly tighter constraints on the neutrino mass. From the 3x2pt data with primary CMB and DESI BAO from the first year of data we obtain
\begin{equation}
    \begin{split}
        \sum m_\nu <& 0.082 \text{eV\ at\ 95\% confidence} \\ &({\rm 3x2pt} + {\rm CMB} + {\rm DESI\ BAO})
    \end{split}
\end{equation}
This represents only a very marginal improvement over constraints from the lensing auto-spectrum only ($\sum m_\nu < 0.083 \text{eV\ at\ 95\% confidence}$ from $C_\ell^{\kappa \kappa}$ + CMB + DESI BAO using the ACT DR6 and \textit{Planck} PR4 lensing power spectra\footnote{Note that this is a slightly weaker constraint than presented in \cite{2024arXiv240403002D} for two reasons. Firstly, we use the \texttt{CamSpec} \textit{Planck} PR4 likelihood as our default high-$\ell$ CMB likelihood, and secondly, the older version of the ACT lensing likelihood used in that work effectively deactivated the lensing norm corrections discussed in Appendix~\ref{app:lklh_corr}.}). The 3x2pt results formally disfavour the minimum mass allowed in the inverted hierarchy ($\min_{\rm inv.} \left[\sum m_\nu\right] = 0.1$eV) at \editA{about 98\% confidence}. \editA{We note that the significantly tighter constraint is again at least in part explained by the DESI preference for a slightly lower value of $\Omega_m$, leading to a larger value of $\sigma_8$ at fixed lensing amplitude, requiring less suppression due to neutrinos.}

\resub{Within the context of $\Lambda {\rm CDM} + \sum m_\nu$ the amplitude of late-time fluctuations is consistent with our findings for $\Lambda$CDM in Sec.\,\ref{subsubsec:structure_growth}. For our fiducial analysis including 3x2pt data alongside primary CMB and BAO we find}
\begin{equation}
    \begin{split}
        \sigma_8 =& 0.816^{+0.01}_{-0.006}\\
        S_8 =& 0.828\pm 0.009 \\ &\left({\rm 3x2pt} + {\rm CMB} + {\rm BAO};\ \Lambda{\rm CDM}+\sum m_\nu\right).
    \end{split}
\end{equation}
 
\subsubsection{Non-flat $\Lambda\rm CDM$} \label{subsubsec:curvature}

Within the standard model of cosmology, the Universe is predicted to be spatially flat as a consequence of inflation. However, primary CMB data from \textit{Planck} has at times shown a preference for a closed universe with the curvature parameter, $\Omega_k<0$ \citep{2018ApJ...864...80O,2021PhRvD.103d1301H,2020NatAs...4..196D}. Since the primary CMB anisotropies alone do not constrain curvature due to a ``geometric degeneracy'' \citep{1988ApJ...325L..17P,1999MNRAS.304...75E} this preference is driven entirely by the lensing-induced peak smearing in \textit{Planck} measurements of the CMB anisotropies which has been shown to be somewhat larger than expected from the measured lensing power spectrum \cite[see e.g.,][]{2016A+A...596A.107P,2020A+A...641A...6P}. This preference is not present in independent analyses of CMB anisotropy measurements from ACT+\textit{WMAP} \citep{2020JCAP...12..047A,2020JCAP...12..045C} and also disappears in combination with BAO. 

\begin{figure}
    \centering
    \includegraphics[width=\linewidth]{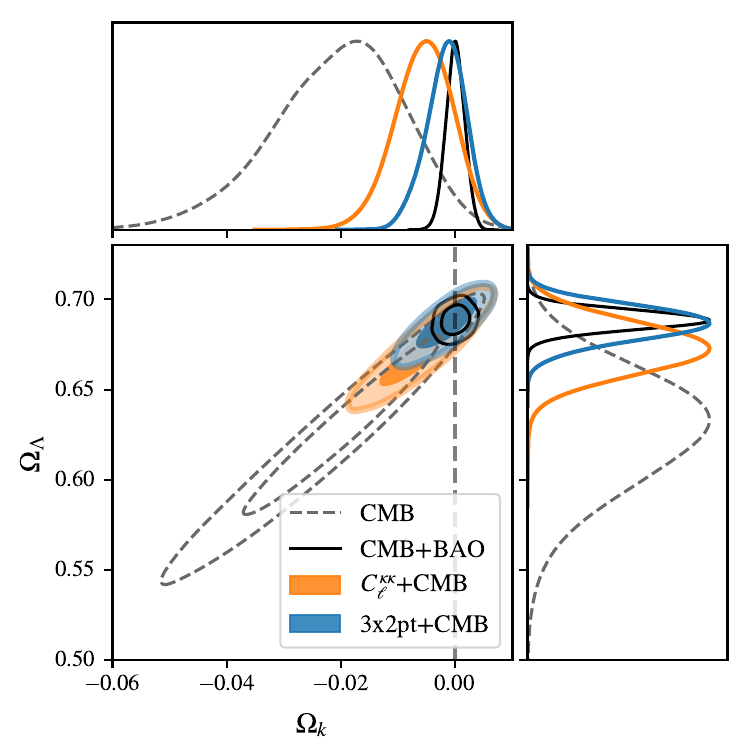}
    \caption{Using a combination of our low redshift lensing data with primary CMB anisotropies we place tight constraints on spatial curvature, showing good consistency with a spatially flat universe. We compare our results to curvature constraints from the primary CMB only, which show a weak preference for a closed universe. When combining primary CMB observations with BAO this preference also disappears and one obtains tight constraints on $\Omega_k$ centred on zero.}
    \label{fig:omk_constraints}
\end{figure}

When combining CMB lensing and unWISE cross-correlation data with primary CMB anisotropies we also no longer observe a preference for a closed universe. We find
\begin{equation}
    \begin{split}
        -0.011 < \Omega_k < 0.004&\text{\ at\ 95\%\ confidence} \\ &({\rm 3x2pt} + {\rm CMB})
    \end{split}
\end{equation}
using both ACT and \textit{Planck} lensing data. We show these constraints in the $\Omega_k$-$\Omega_\Lambda$ plane compared to an analysis of primary CMB anisotropies only as well as the combination of primary CMB anisotropies and BAO in Fig.\,\ref{fig:omk_constraints}. While this constraint represents a significant improvement over the constraint from CMB lensing alone \citep{2024ApJ...962..113M} the combination of primary CMB anisotropies and BAO provides a tighter constraint ($-0.003 < \Omega_k < 0.003\text{\ at\ 95\%\ confidence}$). Despite the fact that the combination of BAO, lensing auto- and cross-spectrum data, and primary CMB does not improve on the constraints from BAO and primary CMB only, our result still represents a valuable cross-check on the BAO derived constraints.

\subsubsection{Extended dark energy models} \label{subsubsec:de_extensions}

\begin{figure}
    \centering
    \includegraphics[width=\linewidth]{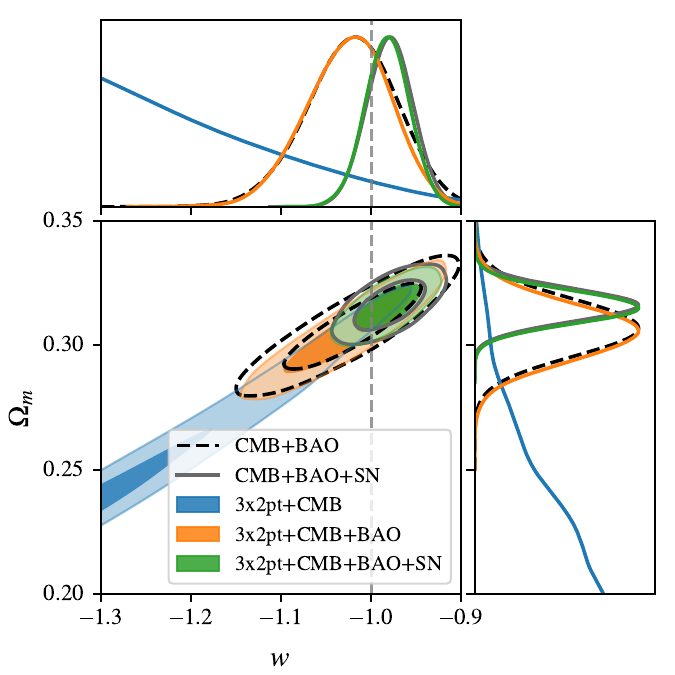}
    \caption{The figure shows constraints on the dark energy equation of state, $w$, and the matter density, $\Omega_m$, in a $w$CDM model. While combining just our 3x2pt data set with primary CMB data is not competitive with CMB + BAO + SN constraints, adding the 3x2pt data to the latter achieves a marginal, 4\%, improvement in constraining power ($w = -0.982\pm 0.025$).}
    \label{fig:w_constraints}
\end{figure}

In this work we consider two extended dark energy scenarios. In the standard $\Lambda$CDM framework dark energy is assumed to be a cosmological constant, $\Lambda$, equivalent to a cosmological fluid with equation of state $w=-1$. First, we consider allowing $w$ to take on values different from $-1$ but remain constant in time. Our data alone is only very weakly sensitive to $w$ and the effect is largely degenerate with other parameters. However, when combining the 3x2pt data set with the primary CMB from \textit{Planck} we find
\begin{equation}
    w = -1.53^{+0.20}_{-0.31}\hspace{0.3cm} ({\rm 3x2pt}+ {\rm CMB})
\end{equation}
with large uncertainties but consistent with $w=-1$ at $\sim$$2\sigma$. This is not competitive with the constraint obtained using all external data sets considered in this work (CMB + BAO + SN) which yields $w = -0.979 \pm 0.026$. Adding the 3x2pt data improves constraints only very marginally by $\sim$4\% to
\begin{equation}
    w = -0.982\pm 0.024\hspace{0.3cm} ({\rm 3x2pt}+ {\rm CMB}+ {\rm BAO}+ {\rm SN}).
\end{equation}
Without supernovae data we find
\begin{equation}
    w = -1.027^{+0.050}_{-0.043}\hspace{0.3cm} ({\rm 3x2pt}+ {\rm CMB}+ {\rm BAO})
\end{equation}
compared to $w = -1.022^{+0.053}_{-0.048}$ from CMB and BAO without our 3x2pt data (a $\sim$7\% improvement). These constraints are shown in Fig.\,\ref{fig:w_constraints}.

We also consider a phenomenological parameterisation of an evolving dark energy equation of state. As in \cite{2023PhRvD.107h3504A} we adopt the CPL model \citep{2001IJMPD..10..213C,2003PhRvL..90i1301L}:
\begin{equation}
    w(a) = w_0 + (1-a) w_a
\end{equation}
where $a=1/(1+z)$ is the scale factor and both $w_0$ and $w_a$ are free parameters. This commonly considered parametrisation has been shown to provide a good fit to several physically motivated dark energy models \citep{2003PhRvL..90i1301L}.

\begin{figure*}
    \centering
    \includegraphics[width=0.5\linewidth]{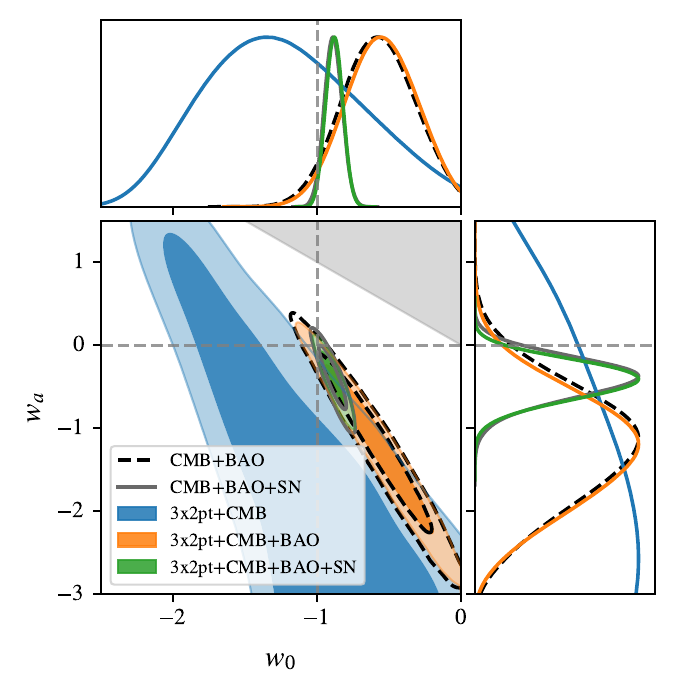}%
    \includegraphics[width=0.5\linewidth]{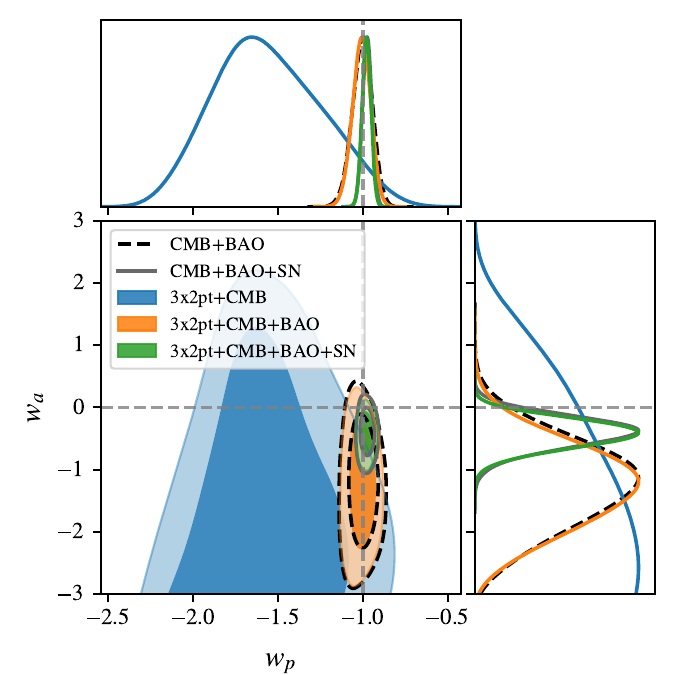}
    \caption{In the \textbf{left} panel we show constraints on dynamical dark energy parametrised by $w(a) = w_0 + (1-a)w_a$. The grey shaded region is excluded by the $w_0 + w_a < 0$ \editA{prior which ensures accelerated expansion}. We find that our data marginally improves constraints on these parameters. Similarly to \cite{2023PhRvD.107h3504A} we determine the redshift at which we obtain the best constraints on the dark energy equation of state within the CPL model. We find this redshift to be $z_p=0.44$ for the combination of CMB and 3x2pt data, $z_p=0.55$ when adding BAO, and $z_p=0.29$ when further adding SN data. This should be compared to $z_p=0.28$ from the external data alone ($z_p=0.54$ from CMB + BAO only). In the \textbf{right} panel we show the constraints on $w_p$ and $w_a$. By construction $w_a$ is non-degenerate with $w_p$ and so we are able to derive meaningful constraints in this parametrisation even for 3x2pt + CMB alone. Note that the pivot redshift and therefore the interpretation of $w_p$ varies between the different data combinations.}
    \label{fig:w0wa_wpwa_constraints}
\end{figure*}

Our constraints on $w_0$ and $w_a$ are shown in Fig.\,\ref{fig:w0wa_wpwa_constraints}. When using only 3x2pt data and the primary CMB $w_0$ and $w_a$ are so significantly degenerate that we are unable to provide meaningful constraints. However, in combination with BAO and SN we find 
\begin{equation}
    \begin{rcases}
        w_0 =& -0.881 \pm 0.060\\
        w_a =& -0.43^{+0.25}_{-0.22}
    \end{rcases}(\text{ 3x2pt + CMB + BAO + SN})
\end{equation}
only a very minor $\sim$5\% and $\sim$9\% improvement on $w_0$ and $w_a$ respectively over constraints from CMB, BAO and SN alone. Removing the supernovae data yields 
\begin{equation}
    \begin{rcases}
        w_0 =& -0.56 \pm 0.24\\
        w_a =& -1.27 \pm 0.66
    \end{rcases}(\text{ 3x2pt + CMB + BAO})
\end{equation}
not improving significantly on constraints from the external data alone (by about 4\% and 3\% respectively). 

To alleviate some of the significant degeneracy between $w_0$ and $w_a$ evident from the left panel in Fig.\,\ref{fig:w0wa_wpwa_constraints} we also determine the `pivot' redshifts, $z_p$, at which we obtain the best constraints on $w(a)$ for various data combinations. Following \cite{2023PhRvD.107h3504A} we estimate $z_p$ by considering the marginalised parameter covariance, defining
\begin{equation}
    a_p = 1 + \frac{C_{w_0 w_a}}{C_{w_a w_a}},
\end{equation} effectively obtaining the best constrained linear combination of $w_0$ and $w_a$ which guarantees that $w_a$ is uncorrelated with $w_p \equiv w(a_p)$. The model can then be reparametrised as $w(a) = w_p + (a_p - a)w_a$. We find $z_p=0.44$, $0.55$, and $0.29$ when combining the 3x2pt data with CMB, CMB + BAO, and CMB + BAO + SN respectively. For comparison the pivot redshift for the external data sets alone (CMB + BAO + SN) is $z_p=0.28$ (or $z_p=0.54$ for CMB + BAO). The equation of state at the pivot redshift is constrained to
\begin{eqnarray}
    w_p &=& -1.56^{+0.28}_{-0.37}\hspace{0.3cm}({\rm 3x2pt} + {\rm CMB}; z_p=0.44),\\
    w_p &=& -1.005^{+0.062}_{-0.055} \hspace{0.3cm}(+ {\rm BAO}; z_p=0.55), \text{\ and}\\
    w_p &=& -0.976 \pm 0.028\hspace{0.3cm}(+ {\rm SN}; z_p=0.29).
\end{eqnarray}
These represent small improvements over constraints from the external data sets alone of about $11\%$, $8\%$ and $7\%$ over CMB, CMB + BAO, and CMB + BAO + SN respectively, in line with the improvements we saw for the $w$CDM model.

We show the two dimensional posterior in the $w_p-w_a$ plane in the right panel of Fig.\,\ref{fig:w0wa_wpwa_constraints}. We can see that all our constraints are consistent with $(w_p, w_a)=(-1, 0)$ within $\sim$$2\sigma$. However, at present the constraining power is insufficient to further shed light on recent hints at deviations from the cosmological constant model with our data similarly consistent with the values preferred by the combination of DESI BAO data and supernovae \cite[$w_0=-0.827\pm 0.063$ and $w_a=-0.75^{+0.29}_{-0.25}$ from CMB + DESI BAO + \texttt{Pantheon+} SN;][]{2024arXiv240403002D}.

\subsubsection{Massive neutrinos in the context of dynamical dark energy} \label{subsubsec:nu_de_extensions}

\begin{figure}
    \centering
    \includegraphics[width=\linewidth]{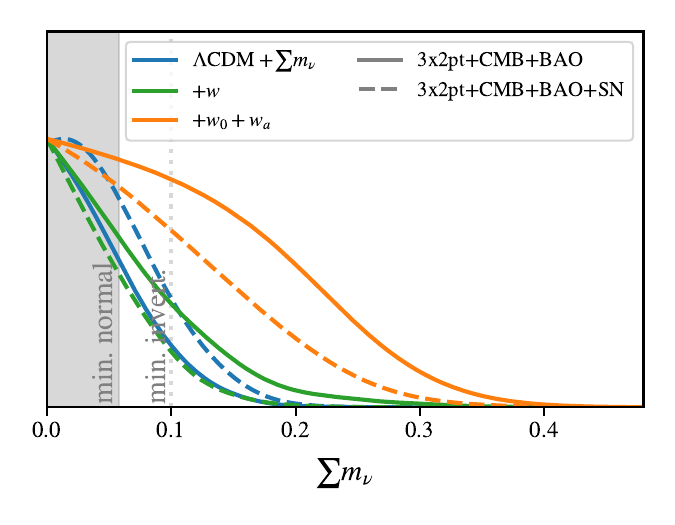}
    \caption{The neutrino mass constraints from large scale structure tracers are sensitive to the cosmological model. In particular the suppression of structure due to neutrino free streaming is degenerate with the impact of modified dark energy. Here we show the one-dimensional marginalised posterior on $\sum m_\nu$ within the $w$CDM+$\sum m_\nu$ and $w_0w_a$CDM+$\sum m_\nu$ models. Due to fluctuations in the preferred central value the one sided constraint on $\sum m_\nu$ within the $w$CDM model is tighter than in the minimally extended $\Lambda$CDM model when including supernovae ($\sum m_\nu < 0.123$eV vs. $\sum m_\nu < 0.137$eV respectively). When considering the $w_0 w_a$CDM model with neutrinos we find significantly degraded constraints ($\sum m_\nu < 0.231$eV when including BAO and SN).}
    \label{fig:mnu_de}
\end{figure}

As has been pointed out several times, the characteristic effect of massive neutrinos on the Universe's background evolution and the suppression of structure formation due to neutrino free streaming can be mimicked by beyond-$\Lambda$CDM extensions \citep[see e.g.,][]{2015PhRvD..92l3535A,2018JCAP...09..017C,2019EPJC...79..262R,2020JCAP...07..037C}. In part because the scale dependence of the neutrino effect is only poorly constrained with current data, the effect is difficult to distinguish from other non-standard physics. Here we consider neutrino mass constraints in the presence of a modified dark energy equation of state\footnote{For constraints on massive neutrinos in a wider range of extended models from the CMB lensing power spectrum alone see \cite{Shao2024}.}. We derive constraints on $\sum m_\nu$, both in the two-parameter extension, $w$CDM+$\sum m_\nu$, as well as in the case of evolving dark energy, $w_0 w_a$CDM+$\sum m_\nu$. Our constraints are summarised in Table \ref{tab:mnu_de}. Additionally, we show the 1D marginalised posteriors on the sum of the neutrino masses in Fig.\,\ref{fig:mnu_de}. 

As expected, we generally see a degradation of constraints in extended dark energy models. One exception occurs when allowing $w$ to vary while including supernovae in the analysis. The supernovae data set provides tight constraints on $w$ with a slight preference for $w>-1$. When $w>-1$ structures grow more slowly and our data are consequently compatible with less suppression due to neutrinos. Within the $w$CDM+$\sum m_\nu$ cosmology our neutrino mass constraint is thus tightened to
\begin{equation}
\begin{split}
    \sum m_\nu <& 0.123 \text{eV\ at\ 95\% c.l.}\\ &({\rm 3x2pt} + {\rm CMB} + {\rm BAO} + {\rm SN})
\end{split}
\end{equation}
compared to $\sum m_\nu < 0.137$eV from the same data in a $\Lambda$CDM+$\sum m_\nu$ model. In the presence of evolving dark energy we find
\begin{equation}
\begin{split}
    \sum m_\nu <& 0.231 \text{eV\ at\ 95\% c.l.}\\ &({\rm 3x2pt} + {\rm CMB} + {\rm BAO} + {\rm SN}).
\end{split}
\end{equation}

%\GSF{do we want to comment on extra tight constraints when restricting to non-phantom dark energy.}

\begin{table*}
\centering
\begin{threeparttable}
    \begin{tabular}{r r | C C C}
    \hline  \hline
        & ~& \sum m_\nu \text{[eV]\ 95\%\ C.I.} & w / w_0 & w_a\\ \hline
        \multirow{2}{*}{$\Lambda$CDM + $\sum m_\nu$} & 3x2pt + CMB + BAO & <0.124 & - & -\\
        &+SN & < 0.137& - & -\\ \hline
        \multirow{2}{*}{$+w$} &3x2pt + CMB + BAO & <0.189 & -1.033^{+0.064}_{-0.047} & -\\
        &+SN & <0.123 & -0.977\pm 0.026 & -\\ \hline
        \multirow{2}{*}{$+w_0+w_a$} &3x2pt + CMB + BAO & <0.273 & -0.53^{+0.26}_{-0.22} \tnote{\dag} & -1.46 \pm 0.71 \tnote{\dag}\\
        &+SN & <0.231 & -0.872^{+0.058}_{-0.066} & -0.50^{+0.34}_{-0.23}\\ \hline \hline
    \end{tabular}
    \begin{tablenotes}
    \item[$\dag$]Constraint affected by prior range
    \end{tablenotes}
    \caption{The characteristic suppression of structure due to neutrino free streaming can be mimicked by extended dark energy models. Here we compare neutrino mass constraints in flat $w$CDM+$\sum m_\nu$ and $w_0w_a$CDM+$\sum m_\nu$ to the ones found for flat $\Lambda$CDM in Sec.\,\ref{subsubsec:nulcdm}. When allowing $w$ to vary but not evolve in time we find tightened constraints when considering supernovae data due to a slight preference for $w>-1$ leading to slower structure growth and therefore tighter bounds on the allowable structure suppression due to neutrinos. In a $w_0w_a$CDM cosmology our neutrino mass constraints are significantly degraded.}
    \label{tab:mnu_de}
\end{threeparttable}
\end{table*}

\section{Conclusion} \label{sec:conclusion}

We have presented cosmological constraints from the joint analysis of the CMB lensing auto-correlation, the cross-correlation between CMB lensing and galaxy data, and the galaxy auto-correlation using CMB lensing data from ACT DR6 and \textit{Planck} PR4 together with galaxy data from unWISE.

Building on previous separate analyses of these data sets we provide tight constraints on the amplitude of matter density fluctuations in the redshift range $z\simeq 0.2-5$. Within the $\Lambda$CDM model we constrain the relevant parameter $S_8^{\rm 3x2pt}\equiv \sigma_8 (\Omega_m/0.3)^{0.4}$ to 1.5\% and in combination with BAO we obtain similarly tight constraints on $\sigma_8$. Our constraints are in excellent agreement with model predictions from $\Lambda$CDM fits to the primary CMB. At the same time our results are not in any statistically significant tension with other LSS probes despite favouring a slightly larger amplitude of matter density fluctuations. Nevertheless, our findings suggest that if the `$S_8$ tension' is indicative of new or so far neglected physics rather than systematic effects, the effect has to be confined either to very low redshifts (probably below $z\simeq 0.5$) where our data are less sensitive or small scales ($k \gtrsim 0.3$ $h$/Mpc) which our data equally does not probe well.

In addition, we constrained the Hubble constant independently of the sound horizon to $3.5\%$. Our constraints are in excellent agreement with constraints from observations of the primary CMB and BAO. While our current uncertainties are too large to provide conclusive evidence, these results raise further questions about the viability of modifications to the sound horizon to resolve the Hubble tension.

We derive constraints on the amplitude of matter density fluctuations as a function of redshift. When the amplitude of low redshift structure is constrained by the cross-correlation data, the lensing auto-spectrum provides constraints on high redshift structures. We find one of the tightest constraints to date ($\sim$$3.3\%$) on the integrated matter density fluctuations above $z\simeq2.4$. The reconstructed $\sigma_8(z)$ is in excellent agreement with predictions from the primary CMB within the standard $\Lambda$CDM model.

Using the 3x2pt data we also revisited constraints on the neutrino mass sum previously presented only from the CMB lensing auto-spectrum. Despite the larger information content we do not tighten existing upper limits. This is because we obtain a slightly smaller amplitude of matter density fluctuations in the joint analysis than in the lensing auto-spectrum analysis, consistent with larger neutrino induced structure suppression and therefore larger neutrino masses. This analysis nevertheless represents a valuable step towards detection of massive neutrinos with future cross- and auto-correlation analyses.

We also investigate, for the first time with this data, a series of beyond-$\Lambda$CDM models including spatial curvature and extended dark energy models. We show that our data marginally tightens some of the existing constraints and provides competitive cross-checks on others. \editA{We also make our data and likelihood publicly available enabling the community to perform further investigation into models not explored in this work (see Appendix~\ref{app:data} for details).}

This work represents the first 3x2pt analysis with the new ACT DR6 lensing reconstruction. Such analyses, combining all possible auto- and cross-correlations between a galaxy sample and the lensing reconstruction, have become the standard in the galaxy weak lensing field. We showed the excellent constraining power of this analysis approach also for CMB lensing data. Similar analyses using future CMB lensing data for example from Simons Observatory \citep[SO;][]{2019JCAP...02..056A} and CMB-S4 \citep{2016arXiv161002743A}, and galaxy samples from future surveys such as Euclid \citep{2011arXiv1110.3193L} and eventually the Vera C. Rubin Observatory's Legacy Survey of Space and Time \citep[VRO LSST;][]{2019ApJ...873..111I} will provide further improved constraints on cosmological parameters and contribute to improving our understanding of the cosmos.

\section*{Acknowledgements}

\input{acknowledgements}

\bibliography{joined_ACT_DR6_lensing,bibliography,software_bib}

\appendix
\onecolumngrid

\section{Data Availability} \label{app:data}

\editA{Pre-release versions of the Markov Chain Monte Carlo runs from this paper and \cite{2024ApJ...966..157F} are available through the NERSC (National Energy Research Scientific Computing Center) Science Gateway, \href{https://portal.nersc.gov/project/act/act_x_unWISE_xcorr+3x2pt}{here}\footnote{\url{https://portal.nersc.gov/project/act/act_x_unWISE_xcorr+3x2pt}}.} 

\editA{Alongside this publication we also distribute various data products and software relevant to this work. The likelihood, bandpowers, covariances and various auxiliary data products required to perform the analysis presented here will be made available upon publication. The likelihood can then be found \href{https://github.com/ACTCollaboration/unWISExLens_lklh}{here}\footnote{\url{https://github.com/ACTCollaboration/unWISExLens_lklh}} and all required data products will be available through the NERSC Science Gateway above and \texttt{\href{https://lambda.gsfc.nasa.gov/product/act/actadv_dr6_lensing_xunwise_info.html}{LAMBDA}\footnote{\url{https://lambda.gsfc.nasa.gov/product/act/actadv_dr6_lensing_xunwise_info.html}}}. We caution that, because of significant correlations, this likelihood should not be combined with any other CMB lensing or CMB lensing cross-correlation likelihoods, like those from \cite{2024ApJ...962..113M} or \cite{2022JCAP...09..039C}. Instead the likelihood provided here can be used to include the auto-correlation measurements in a self consistent manner.}

\section{Likelihood corrections}\label{app:lklh_corr}

The CMB lensing reconstruction is obtained using quadratic estimators which depend on two powers of the observed CMB fields. The normalisation of the estimator and the bias corrections which are required for the lensing power spectrum depend in principle on the underlying CMB power spectra and the lensing convergence power spectrum. In practice we compute the normalisation and bias corrections given a fiducial choice of spectra\footnote{The ACT lensing reconstruction adopts fiducial spectra from a $\Lambda$CDM model fit to \textit{Planck}~2015 TTTEEE data with an updated $\tau$ prior as in~\cite{2017PhRvD..95f3525C}.}. While the CMB power spectrum is well constrained by \textit{Planck}, some residual uncertainty remains that must be propagated to our likelihood analysis.

Let us denote a set of cosmological parameters as $\boldsymbol{\theta}$ and the assumed fiducial cosmology as $\boldsymbol{\theta}_0$. As discussed in more detail in \cite{2024ApJ...962..112Q} the unnormalised lensing reconstruction is sensitive to the product of the lensing convergence field, $\kappa_{L m}(\boldsymbol{\theta})$, and the lensing response function, $\mathcal{R}_L(\boldsymbol{\theta})$. The latter is computed within the fiducial cosmology. When comparing theory to observations we have to account for this fact in the lensing auto- and cross-spectra as
\beq
 C_L^{\kappa g, \rm obs}(\boldsymbol{\theta}) &=& \frac{\mathcal{R}^{-1}_L(\boldsymbol{\theta}_0)}{\mathcal{R}^{-1}_L(\boldsymbol{\theta})} C_L^{\kappa g, \rm th}(\boldsymbol{\theta}), \hspace{0.5cm}\text{and}\\
 C_L^{\kappa \kappa, \rm obs}(\boldsymbol{\theta}) &=& \frac{[\mathcal{R}^{-1}_{L}(\boldsymbol{\theta}_0)]^2}{[\mathcal{R}^{-1}_{L}(\boldsymbol{\theta})]^2}C^{\kappa\kappa}_{L}(\boldsymbol{\theta})-N^1_{L}(\boldsymbol{\theta}_0)+N^1_{L}(\boldsymbol{\theta}).
\eeq
Here we have also included the $N^{1}$-bias in the lensing power spectrum which we compute from simulations and which depends on the fiducial CMB power spectra as well as the lensing power spectrum present in the simulations (see \citealt{2024ApJ...962..112Q} for more details on the $N^{1}$-bias).

Fully calculating the above for each point in the sampled parameter space is unfeasible, and hence we follow the approach of \cite{2016A+A...594A..15P,2020A+A...641A...8P} and \cite{2017PhRvD..95l3529S} and forward model the linearised corrections to the theory spectrum due to the parameter deviations from the fiducial cosmology. Given the excellent constraints on the CMB power spectra any deviations from the fiducial spectrum, $C^{\rm{CMB}}_\ell(\boldsymbol{\theta}_0)$, are expected to be small justifying this expansion. The spectra to be compared to the observed data spectra are thus given by
\beq\label{eq:norm_corr_clkg}
    C_L^{\kappa g, \rm obs}(\boldsymbol{\theta}) &\approx& C_L^{\kappa g, \rm th}(\boldsymbol{\theta})\left[1 - M_L^\ell \left[C^{\rm CMB}_{\ell}(\boldsymbol{\theta})-C^{\rm CMB}_{\ell}(\boldsymbol{\theta}_0)\right]\right], \hspace{0.5cm}\text{and}\\
    C_L^{\kappa \kappa, \rm obs}(\boldsymbol{\theta}) &\approx&
    C^{\kappa\kappa}_{L}(\boldsymbol{\theta})-2M_L^\ell\left[C^{\rm CMB}_{\ell}(\boldsymbol{\theta})-C^{\rm CMB}_{\ell}(\boldsymbol{\theta}_0)\right]C^{\kappa\kappa}_{L}(\boldsymbol{\theta}_0) \nonumber\\ &&+\frac{d{N}^1_{L}}{d{C}^{\rm CMB}_{\ell}}\left[C^{\rm CMB}_{\ell}(\boldsymbol{\theta})-C^{\rm CMB}_{\ell}(\boldsymbol{\theta}_0)\right]+\frac{d{N}^1_{L}}{d{C}^{\kappa\kappa}_{L^{\prime}}}\left[{C}^{\kappa\kappa}_{L^{\prime}}(\boldsymbol{\theta})-{C}^{\kappa\kappa}_{L^{\prime}}(\boldsymbol{\theta}_0)\right] \label{eq:norm_corr_clkk}
\eeq
where $M_{L}^{\ell} = \partial \ln \mathcal{R}^{-1}_L / \partial C_\ell^{\text{CMB}} \rvert_{\boldsymbol{\theta}_0}$ is the linearised normalisation-correction matrix.

For joint constraints with CMB anisotropy spectra, we correct the normalisation and $N^{1}$-bias subtraction at each point in parameter space according to Eqs.\,\ref{eq:norm_corr_clkg} and \ref{eq:norm_corr_clkk}. For cosmology runs that do not include information from the primary CMB, we propagate the uncertainty in the lensing normalisation due to possible fluctuations in the CMB power spectrum into an additional contribution to the covariance. Specifically, we obtain 1000 posterior samples from the ACT DR4 + Planck primary CMB chains and propagate these to the covariance matrix as described in Appendix B of \cite{2024ApJ...962..112Q} and Sec.\,6.1 of \cite{2024ApJ...966..157F}. This step is done consistently to both the ACT and \textit{Planck} parts of the covariance matrix.

\end{document}

%% file: authors_farren.tex
\author{Gerrit~S.~Farren}\email{gfarren@lbl.gov}\affiliation{Physics Division, Lawrence Berkeley National Laboratory, 1 Cyclotron Rd, Berkeley, CA 94720, USA}\affiliation{Berkeley Center for Cosmological Physics, University of California, Berkeley, CA 94720, USA}\affiliation{DAMTP, Centre for Mathematical Sciences, University of Cambridge, Wilberforce Road, Cambridge CB3 OWA, UK}\affiliation{Kavli Institute for Cosmology Cambridge, Madingley Road, Cambridge CB3 0HA, UK}

\author{Alex~Krolewski}
\affiliation{Perimeter Institute for Theoretical Physics, 31 Caroline St. North, Waterloo, ON NL2 2Y5, Canada}
\affiliation{Waterloo Centre for Astrophysics, University of Waterloo, Waterloo, ON N2L 3G1, Canada}

\author{Frank~J.~Qu}
\affiliation{Kavli Institute for Particle Astrophysics and Cosmology, 382 Via Pueblo Mall, Stanford, CA 94305, USA}
\affiliation{SLAC National Accelerator Laboratory, 2575 Sand Hill Road, Menlo Park, CA 94025, USA}
\affiliation{DAMTP, Centre for Mathematical Sciences, University of Cambridge, Wilberforce Road, Cambridge CB3 OWA, UK}\affiliation{Kavli Institute for Cosmology Cambridge, Madingley Road, Cambridge CB3 0HA, UK}

\author{Simone~Ferraro}\affiliation{Physics Division, Lawrence Berkeley National Laboratory, 1 Cyclotron Rd, Berkeley, CA 94720, USA}\affiliation{Berkeley Center for Cosmological Physics, University of California, Berkeley, CA 94720, USA}

%\author{Niall~MacCrann or Frank~J.~Qu (permutation of authors TBD, remaining author to go in alphabetical list)}

%\author{Irene~Abril-Cabezas}\affiliation{DAMTP, Centre for Mathematical Sciences, University of Cambridge, Wilberforce Road, Cambridge CB3 OWA, UK}

%\author{Rui An}\affiliation{Department of Physics and Astronomy, University of Southern California, Los Angeles, CA 90089, USA}

%\author{Zachary~Atkins}\affiliation{Joseph Henry Laboratories of Physics, Jadwin Hall, Princeton University, Princeton, NJ, USA 08544}

%\author{Nicholas~Battaglia}
%\affiliation{Department of Astronomy, Cornell University, Ithaca, NY 14853, USA}

%\author{J.~Richard~Bond}\affiliation{Canadian Institute for Theoretical Astrophysics, University of Toronto, Toronto, ON, Canada M5S 3H8}

\author{Erminia~Calabrese}\affiliation{School of Physics and Astronomy, Cardiff University, The Parade, Cardiff, Wales CF24 3AA, UK}

%\author{Steve~K.~Choi}\affiliation{Department of Physics, Cornell University, Ithaca, NY 14853, USA}\affiliation{Department of Astronomy, Cornell University, Ithaca, NY 14853, USA}

%\author{Omar~Darwish}\affiliation{Universit\'{e} de Gen\`{e}ve, D\'{e}partement de Physique Th\'{e}orique et CAP, 24 quai Ernest-Ansermet, CH-1211 Gen\`{e}ve 4, Switzerland}

%\author{Mark~J.~Devlin}\affiliation{Department of Physics and Astronomy, University of Pennsylvania, 209 South 33rd Street, Philadelphia, PA, USA 19104}

%\author{Adriaan~J.~Duivenvoorden} \affiliation{Center for Computational Astrophysics, Flatiron Institute, New York, NY 10010, USA}\affiliation{Joseph Henry Laboratories of Physics, Jadwin Hall, Princeton University, Princeton, NJ, USA 08544}

\author{Jo~Dunkley}\affiliation{Joseph Henry Laboratories of Physics, Jadwin Hall, Princeton University, Princeton, NJ, USA 08544}\affiliation{Department of Astrophysical Sciences, Peyton Hall, Princeton University, Princeton, NJ 08544, USA}

\author{Carmen~Embil~Villagra}\affiliation{DAMTP, Centre for Mathematical Sciences, University of Cambridge, Wilberforce Road, Cambridge CB3 OWA, UK}

\author{J.~Colin Hill}\affiliation{Department of Physics, Columbia University, 538 West 120th Street, New York, NY 10027, USA}

%\author{Matt~Hilton}\affiliation{Wits Centre for Astrophysics, School of Physics, University of the Witwatersrand, Private Bag 3, 2050, Johannesburg, South Africa}\affiliation{Astrophysics Research Centre, School of Mathematics, Statistics, and Computer Science, University of KwaZulu-Natal, Westville Campus, Durban 4041, South Africa}

%\author{Kevin~M.~Huffenberger}\affiliation{Department of Physics, Florida State University, Tallahassee FL, USA 32306}

\author{Joshua~Kim}\affiliation{Department of Physics and Astronomy, University of Pennsylvania, 209 South 33rd Street, Philadelphia, PA 19104, USA}

%\author{Thibaut~Louis}\affiliation{Universit\'e Paris-Saclay, CNRS/IN2P3, IJCLab, 91405 Orsay, France}

\author{Mathew~S.~Madhavacheril}\affiliation{Department of Physics and Astronomy, University of Pennsylvania, 209 South 33rd Street, Philadelphia, PA 19104, USA}

%\author{Gabriela~A.~Marques}\affiliation{Fermi National Accelerator Laboratory, P. O. Box 500, Batavia, IL 60510, USA}\affiliation{Kavli Institute for Cosmological Physics, University of Chicago, 5640 S. Ellis Ave., Chicago, IL 60637, USA}

%\author{Jeff~McMahon}\affiliation{Department of Astronomy and Astrophysics, University of Chicago, 5640 S. Ellis Ave., Chicago, IL 60637, USA}\affiliation{Kavli Institute for Cosmological Physics, University of Chicago, 5640 S. Ellis Ave., Chicago, IL 60637, USA}\affiliation{Department of Physics, University of Chicago, Chicago, IL 60637, USA}\affiliation{Enrico Fermi Institute, University of Chicago, Chicago, IL 60637, USA}

\author{Kavilan~Moodley}\affiliation{Astrophysics Research Centre, School of Mathematics, Statistics and Computer Science, University of KwaZulu-Natal, Durban 4001, South Africa}

\author{Lyman~A.~Page}\affiliation{Joseph Henry Laboratories of Physics, Jadwin Hall, Princeton University, Princeton, NJ 08544, USA}

\author{Bruce Partridge}\affiliation{Department of Physics and Astronomy, Haverford College, Haverford, PA 19041, USA}

%\author{Emmanuel Schaan}\affiliation{SLAC National Accelerator Laboratory, Menlo Park, CA 94025, USA}\affiliation{Kavli Institute for Particle Astrophysics and Cosmology and Department of Physics, Stanford University, Stanford, CA 94305, USA}

\author{Neelima~Sehgal}\affiliation{Physics and Astronomy Department, Stony Brook University, Stony Brook, NY 11794, USA}

\author{Blake~D.~Sherwin}\affiliation{DAMTP, Centre for Mathematical Sciences, University of Cambridge, Wilberforce Road, Cambridge CB3 OWA, UK}\affiliation{Kavli Institute for Cosmology Cambridge, Madingley Road, Cambridge CB3 0HA, UK}

\author{Crist\'obal Sif\'on}\affiliation{Instituto de F\'isica, Pontificia Universidad Cat\'olica de Valpara\'iso, Casilla 4059, Valpara\'iso, Chile}

\author{Suzanne~T.~Staggs}\affiliation{Joseph Henry Laboratories of Physics, Jadwin Hall, Princeton University, Princeton, NJ 08544, USA}

\author{Alexander~Van~Engelen}\affiliation{School of Earth and Space Exploration, Arizona State University, Tempe, AZ 85287, USA}

%\author{Cristian~Vargas}\affiliation{Instituto de Astrof\'isica and Centro de Astro-Ingenie\'ia, Facultad de F\'isica, Pontificia Universidad Cat\'olica de Chile, Av. Vicu\~na Mackenna 4860, 7820436 Macul, Santiago, Chile}

%\author{Lukas~Wenzl}\affiliation{Department of Astronomy, Cornell University, Ithaca, NY, 14853, USA}

%\author{Martin~White}\affiliation{Department of Physics, University of California, Berkeley, 366 LeConte Hall MC 7300, Berkeley, CA 94720-7300, USA}

\author{Edward~J.~Wollack}\affiliation{NASA/Goddard Space Flight Center, Greenbelt, MD 20771, USA}

%\author{The ACT Collaboration}

%% file: acknowledgements.tex
The authors wish to thank Hironao Miyatake and the HSC team for making HSC chains available to us slightly before their public release. We thank Noah Sailer for useful discussions.

Support for ACT was through the U.S.~National Science Foundation through awards AST-0408698, AST-0965625, and AST-1440226 for the ACT project, as well as awards PHY-0355328, PHY-0855887 and PHY-1214379. Funding was also provided by Princeton University, the University of Pennsylvania, and a Canada Foundation for Innovation (CFI) award to UBC. The development of multichroic detectors and lenses was supported by NASA grants NNX13AE56G and NNX14AB58G. Detector research at NIST was supported by the NIST Innovations in Measurement Science program. 
ACT operated in the Parque Astron\'omico Atacama in northern Chile under the auspices of the Agencia Nacional de Investigaci\'on y Desarrollo (ANID). We thank the Republic of Chile for hosting ACT in the northern Atacama, and the local indigenous Licanantay communities whom we follow in observing and learning from the night sky.

Computing was performed using the Princeton Research Computing resources at Princeton University, the Niagara supercomputer at the SciNet HPC Consortium, and the Symmetry cluster at the Perimeter Institute. This research also used resources provided through the STFC DiRAC Cosmos Consortium and hosted at the Cambridge Service for Data Driven Discovery (CSD3). SciNet is funded by the CFI under the auspices of Compute Canada, the Government of Ontario, the Ontario Research Fund–Research Excellence, and the University of Toronto. Research at Perimeter Institute is supported in part by the Government of Canada through the Department of Innovation, Science and Industry Canada and by the Province of Ontario through the Ministry of Colleges and Universities. This research also used resources of the National Energy Research Scientific Computing Center (NERSC), a U.S. Department of Energy Office of Science User Facility located at Lawrence Berkeley National Laboratory, operated under Contract No. DE-AC02-05CH11231 using NERSC award HEP-ERCAPmp107.

GSF acknowledges support through the Isaac Newton Studentship and the Helen Stone Scholarship at the University of Cambridge. 
GSF, FJQ, CEV, and BDS acknowledge support from the European Research Council (ERC) under the European Union’s Horizon 2020 research and innovation programme (Grant agreement No. 851274). BDS further acknowledges support from an STFC Ernest Rutherford Fellowship. SF is supported by Lawrence Berkeley National Laboratory and the Director, Office of Science, Office of High Energy Physics of the U.S. Department of Energy under Contract No.\ DE-AC02-05CH11231. 
EC acknowledges support from the European Research Council
(ERC) under the European Union’s Horizon 2020 research and innovation programme (Grant agreement No. 849169). CEV received the support from “la Caixa” Foundation (ID 100010434, fellowship code LCF/BQ/EU22/11930099). JK acknowledges support from NSF grants AST-2307727 and AST-2153201. KM acknowledges support from the National Research Foundation of South Africa. MM acknowledges support from NSF grants AST-2307727 and  AST-2153201 and NASA grant 21-ATP21-0145.  CS acknowledges support from the Agencia Nacional de Investigaci\'on y Desarrollo (ANID) through Basal project FB210003.

\subsection*{Software}

%We acknowledge use of the \texttt{matplotlib} \citep{Hunter:2007} package and the Python Image Library for producing plots in this paper. Furthermore, we acknowledge use of the \texttt{numpy} \citep{harris2020array} and \texttt{scipy} \citep{2020SciPy-NMeth} packages. We use the Boltzman code \texttt{CAMB} \citep{Lewis1999,Howlett2012} for calculating theory spectra, and use \texttt{GetDist} \citep{Lewis:2019xzd} and \texttt{Cobaya} \citep{Torrado2021} for likelihood analysis.

We acknowledge use of the \texttt{matplotlib} \citep{2007CSE.....9...90H} package and the Python Image Library for producing plots in this paper. Furthermore, we acknowledge use of the \texttt{numpy} \citep{2020Natur.585..357H} and \texttt{scipy} \citep{2020NatMe..17..261V} packages. We use the Boltzmann code \texttt{CAMB} \citep{2000ApJ...538..473L,2012JCAP...04..027H} for calculating theory spectra, and use \texttt{GetDist} \citep{2019arXiv191013970L} and \texttt{Cobaya} \citep{2021JCAP...05..057T} for likelihood analysis.